\documentclass[twocolumn]{aastex631}
\usepackage{multirow}
\usepackage[graphicx]{realboxes}
\usepackage{amsmath, amssymb}
\usepackage{xcolor}
\definecolor{r2green}{rgb}{0.0,0.5,0.0} 

\shorttitle{EHT Constraints on Einasto Dark Matter in Sgr A$^{*}$ and M87$^{*}$}
\shortauthors{A. Errehymy, S Hansraj and C. Hansraj}

\graphicspath{{./}{figures/}}

\begin{document}

\title{Observational Limits on Einasto Dark Matter Parameters from Event Horizon Telescope Images of Sgr A$^{*}$ and M87$^{*}$}

\author[0000-0002-0253-3578]{A. Errehymy}
\affiliation{Astrophysics Research Centre, School of Mathematics, Statistics and Computer Science, University of KwaZulu-Natal, Private Bag X54001, Durban 4000, South Africa, Email: abdelghani.errehymy@gmail.com}
\affiliation{Center for Theoretical Physics, Khazar University, 41 Mehseti Str., Baku, AZ1096, Azerbaijan}

\author[0000-0002-8305-7015]{S. Hansraj}
\affiliation{Astrophysics Research Centre, School of Mathematics, Statistics and Computer Science, University of KwaZulu-Natal, Private Bag X54001, Durban 4000, South Africa, Email: hansrajs@ukzn.ac.za}

\author[0000-0001-5304-7433]{C. Hansraj}
\affiliation{Applied Mathematics Division, Department of Mathematical Sciences, Stellenbosch University, Private Bag X1, Matieland 7602, South Africa, Email: hansrajc@sun.ac.za}

\begin{abstract}

The Event Horizon Telescope (EHT) has provided images of the supermassive black-holes Sgr A$^{*}$ and M87$^{*}$, enabling direct tests of gravity. Any extended mass-distribution, such as a dark-matter halo, perturbs null-geodesics in the photon-ring regime, making shadow-measurements a probe of inner-halo structure. In this work, we investigate static, spherically-symmetric black-holes surrounded by Einasto-type dark-matter halos and derive constraints from EHT shadow-data. Starting from the Einasto density-profile with parameters ${\varrho_0, \tilde{\alpha}, \tilde{\nu}}$, we construct a metric-function $f(r)=1-2M/r+2M_\infty \tilde{g}(r)$ that interpolates between the black-hole horizon and the asymptotic halo, following the approach of Xu et al. (2018) but adapted specifically to the Einasto scenario. We analyze the photon-potential, null-geodesics, and shadow-radius as functions of black-hole mass $M$, in the non-spinning limit. Using the dimensionless shadow-diameter $d_{\text{sh}}\equiv D\theta/M$ measured by the EHT---$d_{\text{sh}}^{M87*}=11.0\pm1.5$ and $d_{\text{sh}}^{SgrA*}=9.5\pm1.4$---we perform Bayesian parameter-estimation to identify allowed regions in the Einasto parameter-space. Combined with the independently-measured black-hole masses from stellar-dynamics, our results place constraints on the inner dark-matter distribution: for Sgr A$^{*}$, adopting the stellar-orbit mass-prior, we find $\varrho_0 \lesssim 10^{-11},M_\odot/\text{pc}^3$ at $1\sigma$ confidence, while for M87$^{*}$ the bounds are weaker due to distance-uncertainties. The Einasto-index $\tilde{\nu}$ is weakly-constrained, indicating that EHT precision primarily limits the mass enclosed near the photon-sphere rather than the profile-slope. These findings demonstrate that horizon-scale imaging can complement galactic-scale dynamics in probing dark-matter models, including fuzzy-dark-matter scenarios with solitonic cores. Future EHT observations will refine these constraints and distinguish between competing dark-matter descriptions.

\end{abstract}



\section{Introduction}\label{Sec:1}
\setlength{\parindent}{0pt}


The Event Horizon Telescope (EHT) has transformed black-holes from theoretical constructs into observable laboratories for strong-gravity physics. The images of M87$^{*}$ \cite{EventHorizonTelescope:2019dse, EventHorizonTelescope:2019ggy, EventHorizonTelescope:2019uob} and Sagittarius A$^{*}$ \cite{EventHorizonTelescope:2022wkp, EventHorizonTelescope:2022xqj} showed remarkable agreement with Kerr geometry predictions from general relativity, demonstrating the importance of studying black-holes as realistic entities with far-reaching implications for our understanding of high-gravity regimes, including constraints on new physics that complement those derived from primordial nucleosynthesis \cite{Carr:2009jm, Jedamzik:2009uy}. These horizon-scale images enable direct, quantitative comparisons between photon-ring predictions and data, testing both gravity and the near-horizon environment.

These findings also provide a testbed for analyzing modifications to the standard theory of gravity. Clearly any extension to general relativity must have physical consequences that harmonize with EHT data. Quantitative measures have assisted us in placing bounds on phenomena such as black-hole and tidal charges; for instance, EHT shadow data constrain a Reissner--Nordstr\"om-type tidal charge to $|q| \lesssim 0.1\,M^2$ \cite{EventHorizonTelescope:2021dqv, Vagnozzi:2020quf}. It should be mentioned that the observed shadow shapes may also be constructed from non-Kerr models \cite{EventHorizonTelescope:2021dqv, Zakharov:2023lib}; for example, Amir et al. \cite{Amir:2018pcu} demonstrated that Kerr-like wormholes generated similar shadows to Kerr black-holes despite using a metric that modifies Kerr geometry without satisfying gravitational field equations.

A central open question in this area is the distribution of dark matter (DM) near supermassive black-holes (SMBHs). Independent of microphysics, any extended mass component perturbs null geodesics at the percent level in the photon-ring regime, so EHT images offer a novel way to investigate inner-halo structure \cite{Liu:2023oab}. Recently there has been an upsurge in explorations of dark matter halos as well as exotic concepts such as fuzzy dark matter (FDM) \cite{Liu:2023oab, Yuan:2021mzi, Davies:2019wgi, Phoroutan-Mehr:2024cwd}. Among candidates, ultralight/fuzzy dark matter provides wave-like phenomenology on kiloparsec scales and black-hole-adjacent signatures, with reviews mapping the mass window and multi-scale constraints that motivate horizon-scale probes as complementary to galaxy-scale fits \cite{Hui:2016ltb, Ferreira:2020fam, Brito:2015oca}. These proposals would create new geometries near black-holes, and there is considerable interest in determining whether such effects align with EHT data.

In our present work we study shadow profiles in the context of FDM with the help of a suitable density profile nominated a priori. Alternative dark matter models such as FDM, which predict the formation of a dense, stationary solitonic core at the galactic center, can potentially influence the gravitational lensing and dynamics observable in the black-hole's shadow and surrounding emission. The presence of such a dense FDM solitonic core could gravitationally lens the photon ring or alter the accretion flow predicted by the standard $\Lambda$-CDM model or an isolated black-hole. We relate the properties of the FDM wavefunction - specifically the mass and central density of the soliton core - to potential signatures - or lack thereof - in the EHT data. By deriving quantitative constraints from the current non-detection of such features, our work places novel, astrophysically robust limits on the inner dark matter density profile, which can in turn be mapped onto constraints on the FDM particle mass through the solitonic core scaling relation \cite{Schive:2014dra}, complementing those from the Lyman - $\alpha$ forest and stellar kinematics.

On the halo-modeling side, we adopt the Einasto law as our density profile---a standard description of CDM halos with a curvature index that flexibly captures deviations from pure power-law slopes and matches simulations and kinematics better than NFW in many systems \cite{navarro2004inner, merritt2006empirical, gao2008redshift, Dutton:2014xda}. Crucially, it also admits useful analytic forms for projected and lensed quantities \cite{Einasto:1965czb, Retana-Montenegro:2012dbd, Baes:2022pbc}. By utilizing the bounds placed on the deviation of the shadow radius from general relativity, it is possible to constrain the parameter space of these models, effectively ruling out configurations that would produce a shadow size out of harmony with EHT interferometric data \cite{EventHorizonTelescope:2021dqv, Vagnozzi:2020quf}.

We note that while spin-dependent effects---such as superradiance-driven spin-down by ultralight bosons---represent an important avenue for constraining dark matter properties around black-holes, such effects require a rotating (Kerr) spacetime and an axisymmetric matter distribution. Since the present work adopts a static, spherically symmetric framework, spin does not enter the metric or the geodesic calculations, and we defer Kerr+halo extensions to future work. The constraints derived here are therefore valid for the non-spinning limit and should be interpreted accordingly.

The key objectives of this study are as follows: Firstly we consider unified modeling which involves developing a robust BH+Einasto pipeline calibrated to the specific environment of M87$^*$ and transferable to Sgr A$^{*}$. Secondly we examine ULDM integration where we  explicitly map the theoretical predictions of ULDM ``solitonic'' cores onto observables at the event horizon scale. Thirdly, we quantify the degeneracy between the DM halo parameters and the black-hole mass by combining the published EHT shadow diameter with an independent stellar-dynamical mass prior, and identify the regions of Einasto parameter space compatible with observations \cite{Liu:2023oab, EventHorizonTelescope:2024dhe, EventHorizonTelescope:2025dua}.

The paper is organized as follows. In Sect. \ref{Sec:1}, we introduce the main motivation and outline the physical context of the study. Sect. \ref{Sec:2} presents the spacetime geometry and discusses the properties of the Einasto density profile used to describe the dark matter distribution. In Sect. \ref{Sec:3}, we construct the corresponding black-hole metric in the presence of interpolated Einasto dark matter. The associated energy emission rate is examined in Sect. \ref{Sec:4}. Sect. \ref{Sec:5} focuses on the geodesic trajectories of particles and photons moving in this spacetime, while Sect. \ref{Sec:6} investigates the properties of circular motion. The stability of null geodesics is then analyzed in Sect. \ref{Sec:7} within a dynamical systems framework. In Sect. \ref{Sec:8}, we compare the theoretical predictions with observational constraints derived from shadow measurements by the Event Horizon Telescope. Finally, Sect. \ref{Sec:9} summarizes the main findings and provides the concluding remarks.

\section{Spacetime Geometry and the Einasto Density Profile}\label{Sec:2}

To describe the gravitational environment of M87$^{*}$, we adopt a static, spherically symmetric line element:
\begin{equation}\mathrm{d}s^2 = -A(r)\mathrm{d}t^2 + B(r)\mathrm{d}r^2 + r^2\mathrm{d}\Omega^2\label{eq:metric}
\end{equation}
where $d\Omega^2 = d\theta^2 + \sin^2 \theta d\phi^2$.
In this framework, the metric functions $A(r)$ and $B(r)$ incorporate the combined gravitational influence of a central black-hole with mass $M_\bullet$ and a surrounding dark matter halo characterized by an Einasto density distribution. This setup aligns with recent high-resolution modeling of M87$^*$, facilitating a direct mapping between the Einasto parameters $(\varrho_e, r_e, \alpha)$ and the resulting spacetime curvature near the photon sphere and its associated critical impact parameter \cite{Liu:2023oab}. For any metric exhibiting these symmetries, the radius of the photon sphere, $r_{\mathrm{ph}}$, and the critical impact parameter, $b_c$, are determined by the null geodesic conditions:
\begin{equation}
r_{\mathrm{ph}}\,A'(r_{\mathrm{ph}}) - 2A(r_{\mathrm{ph}}) = 0, \quad b_c = \frac{r_{\mathrm{ph}}}{\sqrt{A(r_{\mathrm{ph}})}}.\label{eq:photon-sphere}
\end{equation}
When observed from a distance $D$, the angular diameter of the resulting emission ring is given by:
\begin{equation}
\theta_{\mathrm{model}} = \frac{2\,b_c}{D}.\label{eq:ring-diameter}
\end{equation}
While secondary factors such as the specific morphology of the accretion flow can introduce subtle variations in the observed diameter, we account for these uncertainties through marginalization during the parameter estimation process \cite{EventHorizonTelescope:2024dhe, EventHorizonTelescope:2025dua}.We evaluate Eq.~\eqref{eq:ring-diameter} against the M87$^{*}$ observations provided by the EHT, which have demonstrated a remarkably stable ring diameter across the 2017 and 2018 observing runs. This persistence justifies the use of a Gaussian likelihood function:
\begin{equation}
\mathcal{L}{\mathrm{EHT}}(\boldsymbol{\vartheta}) \propto \exp\left[-\frac{\big(\theta{\mathrm{model}}(\boldsymbol{\vartheta})-\theta_{\mathrm{EHT}}\big)^2}{2\,\sigma_\theta^2}\right],\label{eq:likelihood}
\end{equation}
where $\boldsymbol{\vartheta} = (\varrho_e, r_e, \alpha, \ldots)$ represents our parameter space. The values $(\theta_{\mathrm{EHT}}, \sigma_\theta)$ are derived from the EHT's published posterior distributions for the ring size \cite{EventHorizonTelescope:2024dhe, EventHorizonTelescope:2025dua}. For robustness, this methodology can be mirrored using Sgr A$^{*}$ data as an independent secondary check \cite{EventHorizonTelescope:2022wkp}. This Bayesian approach allows us to infer the Einasto parameters $(\varrho_e, r_e, \alpha)$ by combining large-scale galactic dynamics and N-body simulation priors with horizon-scale observational constraints. By contrasting the Einasto results with traditional cusp or spike models, we can better quantify how sensitive the shadow size is to the slope of the inner DM profile. The resulting constraints offer insights into the nature of dark matter---spanning Cold Dark Matter (CDM) and Ultralight Dark Matter (ULDM) paradigms.

\subsection{The Einasto halo model}

We model the DM distribution with the Einasto density profile given by
\begin{equation}
  \varrho(r)=\varrho_e\;\exp\!\left[-\frac{2}{\alpha}\left(\left(\frac{r}{r_e}\right)^{\alpha}-1\right)\right],
  \label{eq:einasto}
\end{equation}
with parameters $(\varrho_e,r_e,\alpha)$. The enclosed mass is
\begin{equation}
  M(<r)=4\pi \varrho_e r_e^3\,\frac{\alpha^{3/\alpha}}{2^{3/\alpha}}\,
  \gamma\!\left(\frac{3}{\alpha},\,\frac{2}{\alpha}\left(\frac{r}{r_e}\right)^{\alpha}\right),
  \label{eq:einasto-mass}
\end{equation}
where $\gamma$ is the lower incomplete gamma function \cite{Retana-Montenegro:2012dbd}.

The Einasto profile has emerged as a flexible and observationally successful description of galactic dark matter distributions since its introduction in studies of the Andromeda galaxy and the Milky Way \cite{einasto1965kinematics, einasto1969galactic, einasto1969andromeda}. In this formulation, the density profile $\varrho(r)$ serves as the fundamental entity from which all other galactic properties - such as gravitational potentials, cumulative mass, and dynamical signatures, can be derived. For a density profile to represent a physically plausible galactic system, it must satisfy several basic regularity conditions: smoothness across all radii $\varrho(r)\in C^{\infty}(\mathbb{R}^{+})$, asymptotic vanishing at large distances $\lim_{r\rightarrow\infty}\varrho(r)=0$, and strict positivity with no singularities $0<\varrho(r)<\infty$  for all $r > 0$). Moreover, integrated quantities like total mass, half-mass radius, and central potential must remain finite. These criteria ensure that models built upon the Einasto proposal remain both mathematically well - behaved and astrophysically meaningful when applied to real galaxies.

Over the past two decades, the Einasto model has become a standard tool in cold dark matter simulations \cite{navarro2004inner,springel2005simulations,mamon2005dark,hayashi2008understanding,gao2008redshift} and has proven particularly useful for describing halos spanning a wide mass range---from dwarf spheroidals \cite{graham2003hst} to massive clusters \cite{hayashi2008understanding}. It has also been employed to estimate local dark matter densities from rotation curves \cite{de2019estimation} and to characterize bulge and bar structures in SDSS galaxies \cite{gadotti2009structural}. Its key advantage lies in the Einasto index, which controls the curvature of the profile and enables a more accurate representation of simulated halo shapes than the standard Navarro-Frenk-White (NFW) formula \cite{merritt2006empirical}. This flexibility motivates our choice of the Einasto profile as the foundation for modeling dark matter environments around supermassive black-holes.

Beyond its application to individual galaxies such as Sculptor dwarfs, M31, M32, M87, and the Milky Way \cite{einasto1969galactic,einasto1969andromeda}, the Einasto dark matter (DM) model has proven valuable for describing the density profiles of DM haloes \cite{navarro2004inner,springel2005simulations,mamon2005dark,hayashi2008understanding,gao2008redshift}. Analytical investigations inspired by the Einasto profile have explored spherically symmetric galaxy models, including both spiral and analytical types, as well as their associated DM haloes, with a particular focus on characterizing the logarithmic slope \cite{cardone2005spherical,dhar2010surface,Retana-Montenegro:2012dbd}. The model is widely used in simulations of $\Lambda$ cold dark matter haloes \cite{gao2008redshift,hayashi2008understanding,merritt2006empirical}, in estimating local DM density from galactic rotation curves \cite{de2019estimation}, and in examining structural features of spiral and elliptical galaxies, including bulges and bars, as revealed by SDSS surveys \cite{gadotti2009structural}. Moreover, it has been applied to study the properties of dwarf elliptical galaxies \cite{graham2003hst}, highlighting its versatility in capturing the complex distribution of dark matter across a wide range of galactic systems.

The distribution of matter can also be described by the Einasto profile \cite{einasto1969galactic,einasto1969andromeda}, given by
\begin{align}\label{s1}
\varrho(r)=\varrho_{s}\exp\left\{-d_{\tilde{\nu}}\left[\left(\frac{r}{r_{s}}\right)^{1/\tilde{\nu}}-1\right]\right\},
\end{align}
in which $r_{s}$ represents the characteristic (half-mass) radius, $d_{\tilde{\nu}}$ is a shape-dependent constant fixed by the choice of $r_{s}$, $\varrho_{s}$ specifies the density at this scale, and $\tilde{\nu}$ is the Einasto index that governs the curvature of the profile. Several parameterizations exist for the Einasto density profile in the literature. One common form for dark matter halos is
\begin{align}\label{s2}
\varrho(r)=\varrho_{-2}\exp\left\{-2\tilde{\nu}\left[\left(\frac{r}{r_{-2}}\right)^{1/\tilde{\nu}}-1\right]\right\},
\end{align}
where $r_{-2}$ and $\varrho_{-2}$ correspond to the radius and density at which the logarithmic slope satisfies $\frac{d\ln \varrho}{d\ln r}=-2$.

Introducing the central density and rescalings 
\begin{align}\label{s3}
\varrho_{0}=\varrho_{s}\exp\left[d_{\tilde{\nu}}\right]=\varrho_{-2}\exp\left[2\tilde{\nu}\right], \quad
\tilde{\alpha} = \frac{r_{-2}}{2^{\tilde{\nu}} \tilde{\nu}^{\tilde{\nu}}} = \frac{r_s}{d_{\tilde{\nu}}^{\tilde{\nu}}}
\end{align}
the profile can be compactly expressed as
\begin{align}\label{s5}
\varrho(r)=\varrho_{0}\exp\left[-\left(\frac{r}{\tilde{\alpha}}\right)^{{1/\tilde{\nu}}}\right].
\end{align}

The flexibility of the Einasto profile arises from its parameters $\{\varrho_{0}, \tilde{\nu}, \tilde{\alpha}\}$, which allow it to accommodate halos ranging from dwarf galaxies to massive clusters. For dark matter halos spanning this mass range, the Einasto index typically lies between $4.54 \lesssim \tilde{\nu} \lesssim 8.33$, with an average value of $\tilde{\nu} \simeq 5.88$ \cite{navarro2004inner}. Analyses indicate that $\tilde{\nu}$ decreases with increasing redshift and halo mass, taking values around $\tilde{\nu} \sim 4.35$ for cluster-sized halos and $\tilde{\nu} \sim 5.88$ for galaxy-sized halos in the Millennium Run \cite{springel2005simulations}. Similar patterns have been observed in galaxy-sized halos within the Aquarius simulations \cite{springel2008aquarius}.

\begin{figure*}
\centering
\includegraphics[width=4.5cm,height=4.5cm]{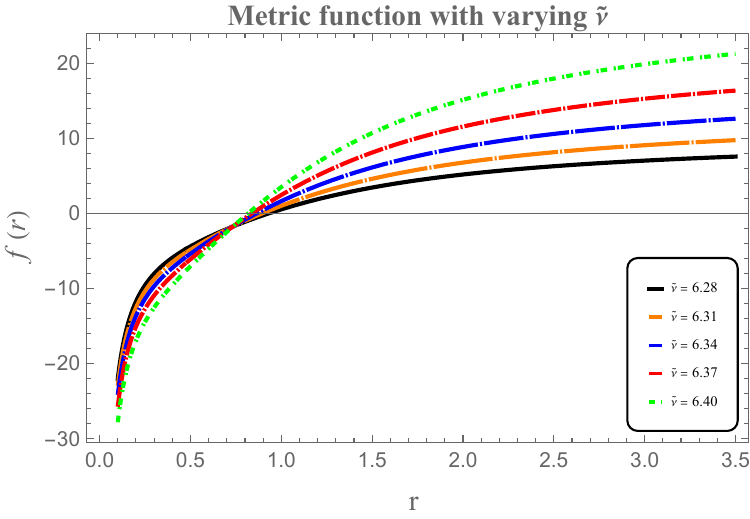}%
\includegraphics[width=4.5cm,height=4.5cm]{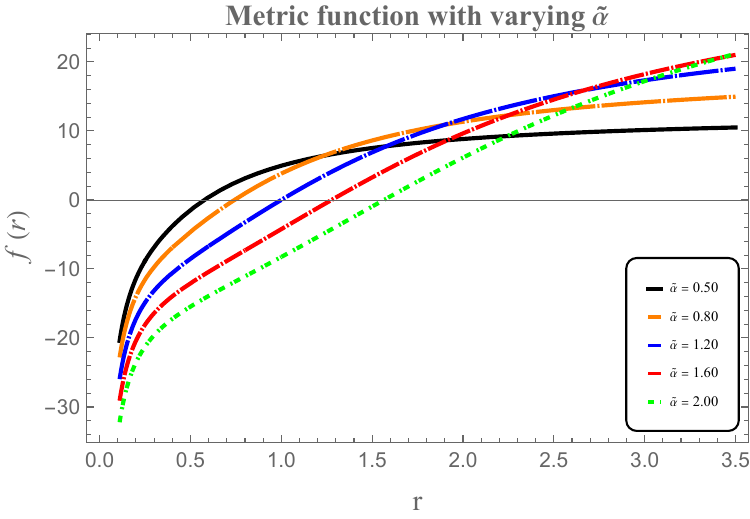}%
\includegraphics[width=4.5cm,height=4.5cm]{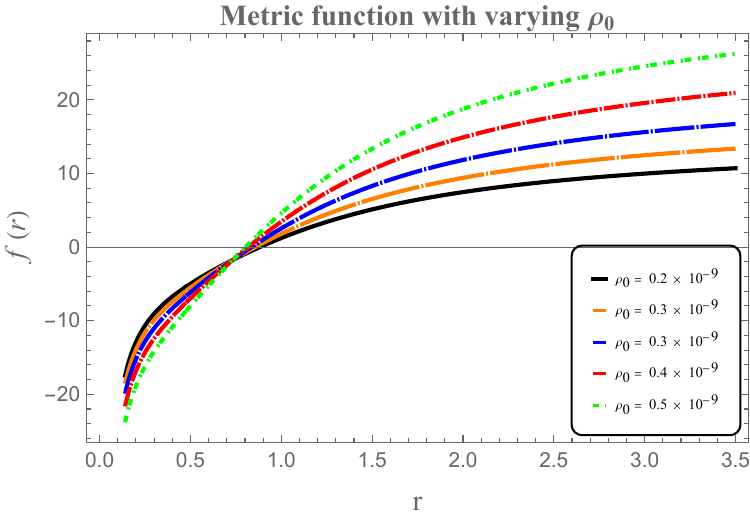}%
\includegraphics[width=4.5cm,height=4.5cm]{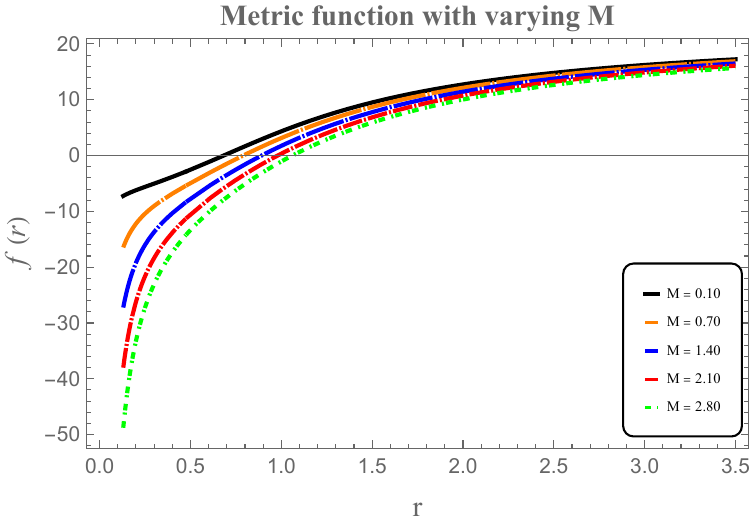}%
    \caption{\footnotesize The lapse function $f(r)$ is analyzed in the presence of an Einasto dark matter distribution, with its behavior governed by the parameters $\tilde{\nu}$, $\tilde{\alpha}$, $\varrho_0$, and the black-hole mass $M$. The leftmost panel illustrates the effect of varying $\tilde{\nu}=6.28, 6.31, 6.34, 6.37,$ and $6.40$, while keeping $\tilde{\alpha}=0.95$, $\varrho_0=10^{-16.75}$, and $M=1.0$ fixed. The second panel from the left shows how changes in $\tilde{\alpha}=0.5, 0.8, 1.2, 1.6,$ and $2.0$ modify the profile, with $\tilde{\nu}=4.39$, $\varrho_0=10^{-9.5}$, and $M=1.0$ held constant. The third panel examines the impact of varying the dark matter density parameter $\varrho_0=10^{-9.7}, 10^{-9.6}, 10^{-9.5}, 10^{-9.4},$ and $10^{-9.3}$, while fixing $\tilde{\nu}=4.39$, $\tilde{\alpha}=0.95$, and $M=1.0$. The rightmost panel highlights the influence of the black-hole mass $M=0.1, 0.7, 1.4, 2.1,$ and $2.8$ on the lapse function, with $\tilde{\nu}=4.39$, $\tilde{\alpha}=0.95$, and $\varrho_0=10^{-9.5}$ kept unchanged. All quantities are in geometric units ($G = c = 1$); $r$ is measured in units of $M$ and the mass values $M = 0.1, \ldots, 2.8$ are dimensionless ratios.} \label{Fig1}
\end{figure*}

\section{black-hole metric with interpolated Einasto dark matter}\label{Sec:3}

\noindent\textit{Note on units.}---Throughout Sections~\ref{Sec:3}--\ref{Sec:7}, we work in geometric units with $G = c = 1$, so that all lengths are expressed in units of the black-hole mass $M$ (i.e.\ the gravitational radius $r_g = GM/c^2$). For reference, $1\,M \approx 6.4\times 10^{6}\,$km for Sgr\,A$^*$ and $1\,M \approx 9.6\times 10^{9}\,$km for M87$^*$. The parameters $\varrho_0$, $\tilde{\alpha}$, and $\tilde{\nu}$ appearing in the figures are likewise dimensionless in these units unless otherwise stated.

\medskip

We now introduce the function
\begin{align}\label{s6}
\gamma(\alpha,x)=\int_{0}^{x}\exp[-z] z^{\alpha-1} dz\quad (\text{Re}~\alpha>0),
\end{align}
known as the lower incomplete Gamma function. This definition naturally arises from decomposing the complete Gamma function \cite{prym1877theorie},
\begin{align}
\Gamma(\alpha)=\int_{0}^{\infty}\exp[-z]z^{\alpha-1}dz.
\end{align}

The exact enclosed mass for an Einasto halo involves an integral over the density profile, which evaluates to 
\begin{align}
M(r) = 4\pi \int_0^r \varrho(\tilde r) \, \tilde r^2 \, d\tilde r = 4\pi \varrho_0 \tilde{\alpha}^3 \tilde{\nu} \, \gamma\Big(3\tilde{\nu}, (r/\tilde{\alpha})^{1/\tilde{\nu}}\Big),
\end{align}
where $\gamma(\alpha,x)$ denotes the lower incomplete Gamma function. Although this expression is formally exact, its dependence on the incomplete Gamma function makes fully analytic manipulation cumbersome - particularly when the metric function must be integrated further to obtain photon orbits and shadow radii. To maintain analytical tractability while preserving the essential physics, we introduce a simplified mass function that captures the asymptotic behavior of the exact profile:
\begin{align}
M(r) \approx M_\infty \frac{r^3}{r^3 + \tilde{\alpha}^3},
\end{align}
with $M_\infty = 4\pi \varrho_0 \tilde{\alpha}^3 \tilde{\nu} \Gamma(3\tilde{\nu})$ being the total halo mass.  This Pad\'e-type approximation reproduces the two limiting regimes of a realistic halo: at small radii   
($r \ll \tilde{\alpha}$), $M(r) \sim M_\infty (r/\tilde{\alpha})^3$)
$(r \ll \tilde{\alpha})$, the enclosed mass scales as $r^3$, consistent with a finite central density and no cusp; at large radii $(r \gg \tilde{\alpha})$, the mass approaches the total halo mass $M_{\infty}$, ensuring a smooth transition to the exterior vacuum. Between these limits, the function interpolates monotonically without introducing unphysical features such as local extrema or discontinuities. This approach follows the strategy employed by Xu et al. \cite{Xu:2018wow} and subsequent studies \cite{Jusufi:2020cpn, Hou:2018bar} for modeling black-holes embedded in extended matter distributions, but here we tailor it specifically to the three-parameter Einasto family. The approximation remains accurate to within a few percent across the radial range relevant for photon-ring calculations (roughly few to tens of gravitational radii), which is sufficient given current observational uncertainties from the EHT.

The tangential (circular) velocity $v_T(r)$ of a test particle in a circular orbit at radius $r$ within the halo is defined via $v_T^2(r) = r\,\Phi'(r)$, where $\Phi(r)$ is the Newtonian gravitational potential. For a spherically symmetric mass distribution this reduces to $v_T^2(r) = M(r)/r$. Thus,
\begin{align}
v_T^2(r) = \frac{M(r)}{r} \approx M_\infty \frac{r^2}{r^3 + \tilde{\alpha}^3},
\end{align}
which naturally interpolates between solid-body rotation near the core and the familiar Keplerian decline at large radii.

To describe the spacetime geometry, we consider a static, spherically symmetric metric:
\begin{align}
ds^2 = -g(r) \, dt^2 + g(r)^{-1} \, dr^2 + r^2 \big(d\theta^2 + \sin^2\theta \, d\phi^2 \big),
\end{align}
where the metric function $g(r)$ is related to the tangential velocity via
\begin{align}
v_T^2(r) = \frac{r}{2} \frac{g'(r)}{g(r)} \quad \Rightarrow \quad g(r) = g_0 \exp\Big[ 2 \int \frac{v_T^2(r)}{r} \, dr \Big],
\end{align}
(note that the factor of $\tfrac{1}{2}$ in the first relation arises from the relativistic circular orbit condition in the static metric, distinguishing it from the Newtonian expression $v_T^2 = M(r)/r$), and $g_0$ is fixed by an appropriate boundary condition, such as $g_0\equiv g(r \to \infty) = 1$ to ensure asymptotic flatness. Substituting the interpolated velocity yields
\begin{align}
g(r) = \exp\Big[ 2 M_\infty \int \frac{r}{r^3 + \tilde{\alpha}^3} \, dr \Big].
\end{align}

\begin{figure*}
\centering
\includegraphics[width=4.5cm,height=4.5cm]{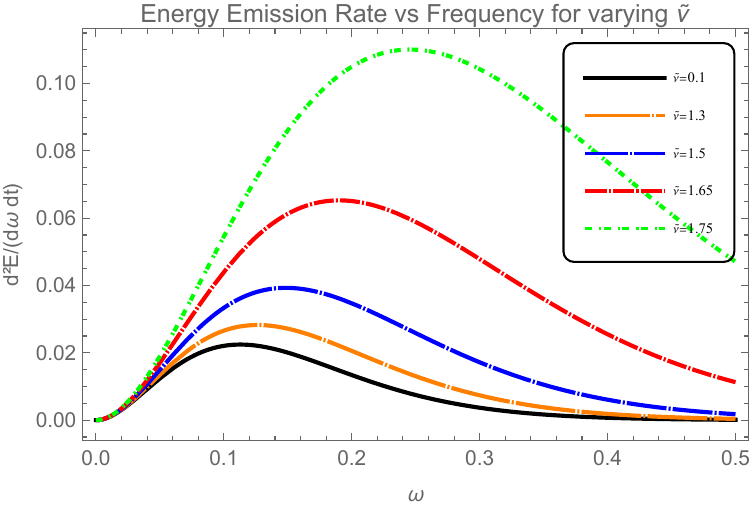}%
\includegraphics[width=4.5cm,height=4.5cm]{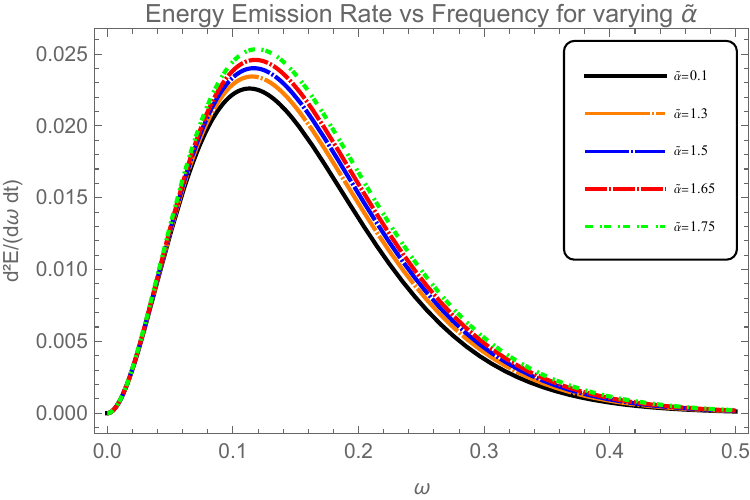}%
\includegraphics[width=4.5cm,height=4.5cm]{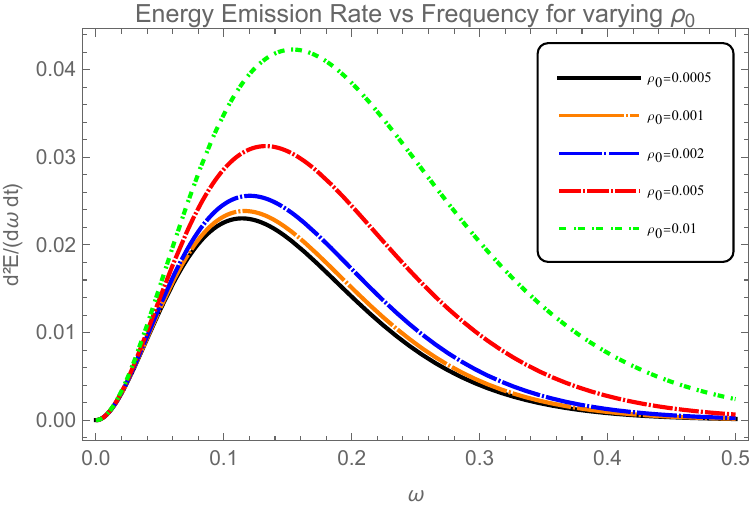}%
\includegraphics[width=4.5cm,height=4.5cm]{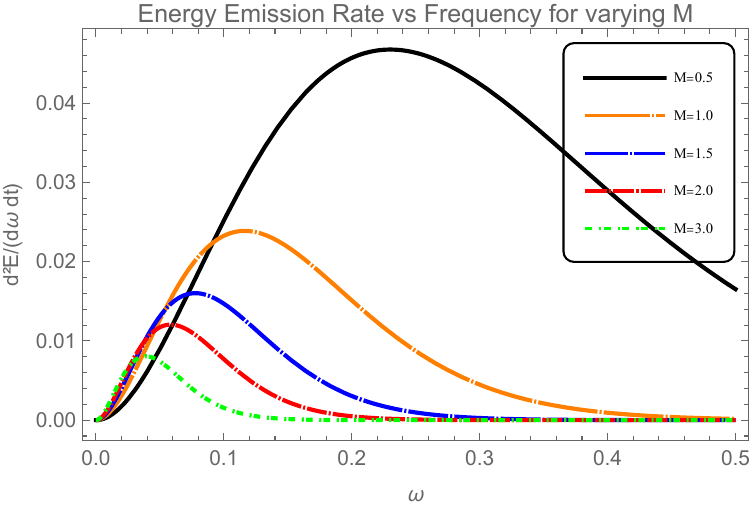}%
    \caption{\footnotesize The energy emission rate $\frac{d^2 E}{d\omega \, dt}$ is examined in the framework of an Einasto dark matter distribution, where its behavior is regulated by the parameters $\tilde{\nu}$, $\tilde{\alpha}$, $\varrho_0$, and the black-hole mass $M$. The leftmost panel displays the response of the emission rate to variations in $\tilde{\nu}=0.1, 1.3, 1.5, 1.65,$ and $1.75$, while $\tilde{\alpha}=0.5$, $\varrho_0=10^{-3}$, and $M=1.0$ are kept fixed. The second panel from the left illustrates the influence of changing $\tilde{\alpha}=0.1, 0.35, 0.55, 0.75,$ and $1.0$, with $\tilde{\nu}=1.0$, $\varrho_0=10^{-3}$, and $M=1.0$ held constant. The third panel highlights the effect of varying the dark matter density parameter $\varrho_0=0.0005, 0.001, 0.002, 0.005,$ and $0.01$, while maintaining $\tilde{\nu}=1.0$, $\tilde{\alpha}=0.5$, and $M=1.0$. The rightmost panel demonstrates how different values of the black-hole mass $M=0.5, 1.0, 1.5, 2.0,$ and $3.0$ modify the energy emission rate, with $\tilde{\nu}=1.0$, $\tilde{\alpha}=0.5$, and $\varrho_0=10^{-3}$ remaining fixed. All quantities are in geometric units ($G = c = \hbar = 1$): the frequency $\omega$ is in units of $M^{-1}$ and the emission rate $d^2E/(d\omega\,dt)$ is dimensionless.} \label{Fig1a}
\end{figure*}

This integral can be evaluated analytically:
\begin{align}
\int \frac{r}{r^3 + \tilde{\alpha}^3} \, dr = \tilde{g}(r) + C,
\end{align}
where 
\begin{align}
\tilde{g}(r)=\frac{1}{6\tilde{\alpha}^2} \ln\left| \frac{r^2 - r \tilde{\alpha} + \tilde{\alpha}^2}{r^2 + r \tilde{\alpha} + \tilde{\alpha}^2} \right| + \frac{1}{\sqrt{3} \tilde{\alpha}^2} \arctan\left( \frac{2r - \tilde{\alpha}}{\sqrt{3} \tilde{\alpha}} \right),
\end{align}
leading to a fully analytical expression for the metric function:
\begin{align}
g(r) =e^{2 M_\infty \tilde{g}(r)}\approx 1+ 2 M_\infty \tilde{g}(r).
\end{align}

This form is smooth, differentiable, and positive for all $r > 0$, well-behaved at the origin, and reduces to the Schwarzschild metric in the asymptotic regime ($g(r) \sim 1 - 2 M_\infty / r$). The approach provides a tractable framework for analyzing black-holes embedded within dark matter halos. Building on the formulation introduced by Xu et al. \cite{Xu:2018wow} and later developed in related works \cite{Jusufi:2020cpn, Hou:2018bar, Xu:2021dkv}, the spacetime under consideration represents a static and spherically symmetric black-hole encompassed by a dark matter halo. The surrounding matter follows the Einasto-type density distribution, which is carefully normalized to remain consistent and physically meaningful across all radial distances.
\begin{align} 
ds^{2}=-f\left( r\right) dt^{2}+\frac{1}{f\left( r\right) }dr^{2}+r^{2}\left(
d\theta ^{2}+\sin ^{2}\theta d\phi ^{2}\right) ,\label{metric1}
\end{align}
where
\begin{align}\label{laps1}
f\left( r\right) = 1-\frac{2M}{r} + 2 M_\infty \tilde{g}(r).~
\end{align}%
Here, the first term is the familiar Schwarzschild contribution, while the second term accounts for the dark matter distributed according to the Einasto profile. The location of the event horizon, $r_+$, is determined by solving
\begin{align}
f(r_+) = 0,  
\end{align}
which balances the pull of the central mass against the dark matter contribution. We can explore the BH solution's behavior by taking advantage of curvature scalars. The Kretschmann scalar for the metric (\ref{metric1}) is
\begin{align}
K(r) &= R^{\mu\nu\alpha\beta} R_{\mu\nu\alpha\beta} =
 \bigl(-\frac{4 M}{r^3} + 2 M_\infty \tilde{g}''(r)\bigr)^2 \nonumber\\&+ \frac{4}{r^2} \bigl(\frac{2 M}{r^2} + 2 M_\infty \tilde{g}'(r)\bigr)^2 + \frac{4}{r^4} \bigl(-\frac{2 M}{r} + 2 M_\infty \tilde{g}(r)\bigr)^2
\end{align}
with
\begin{align}
\tilde{g}'(r) &= \frac{1}{6 \tilde{\alpha}^2} \Biggl[ \frac{2 r - \tilde{\alpha}}{r^2 - r \tilde{\alpha} + \tilde{\alpha}^2} - \frac{2 r + \tilde{\alpha}}{r^2 + r \tilde{\alpha} + \tilde{\alpha}^2} \Biggr] \nonumber\\&+ \frac{3}{2 \sqrt{3} \tilde{\alpha} (r^2 - r \tilde{\alpha} + \tilde{\alpha}^2)},\nonumber\\
\tilde{g}''(r) &= \frac{1}{6 \tilde{\alpha}^2} \Biggl[ \frac{-2 r (r - \tilde{\alpha})}{(r^2 - r \tilde{\alpha} + \tilde{\alpha}^2)^2} - \frac{-2 r (r + \tilde{\alpha})}{(r^2 + r \tilde{\alpha} + \tilde{\alpha}^2)^2} \Biggr] \nonumber\\& - \frac{3 (2 r - \tilde{\alpha})}{2 \sqrt{3} \tilde{\alpha} (r^2 - r \tilde{\alpha} + \tilde{\alpha}^2)^2}.\nonumber
\end{align}

\begin{figure*}
\begin{center}
\includegraphics[width=4.5cm,height=4.5cm]{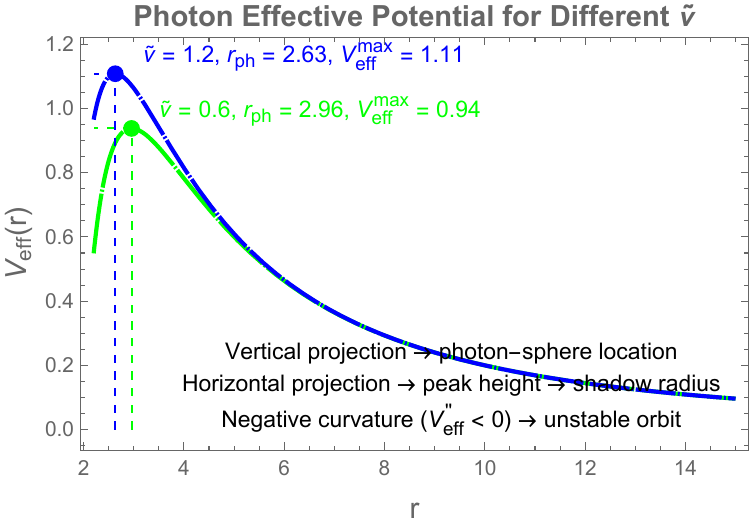}%
\includegraphics[width=4.5cm,height=4.5cm]{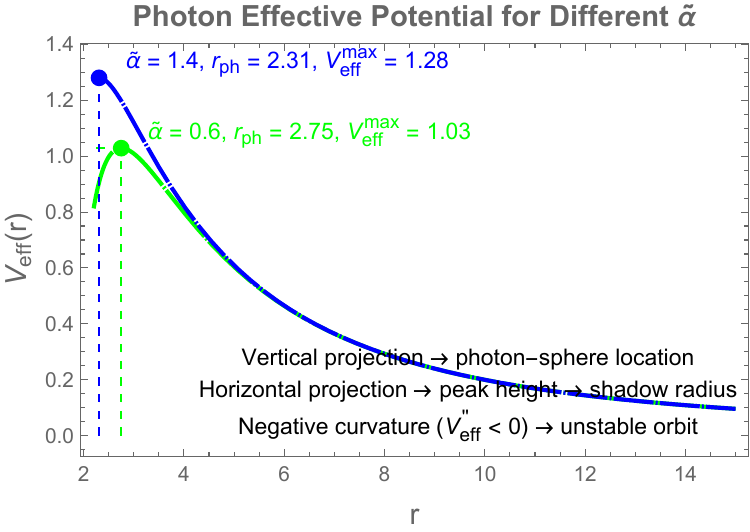}%
\includegraphics[width=4.5cm,height=4.5cm]{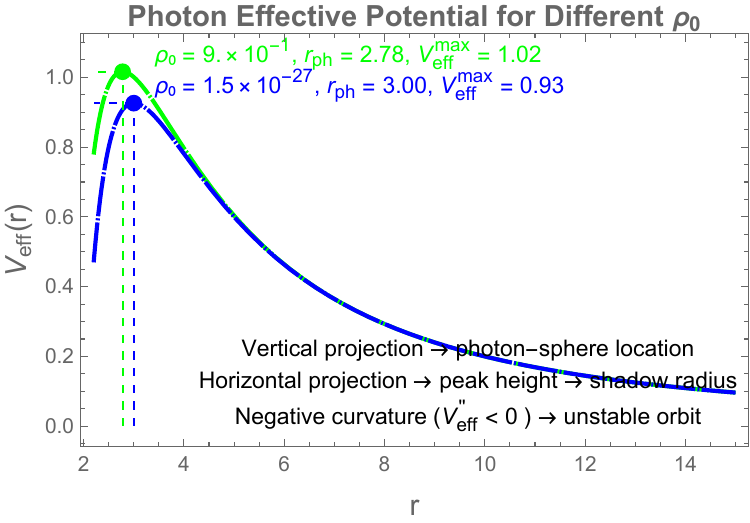}%
\includegraphics[width=4.5cm,height=4.5cm]{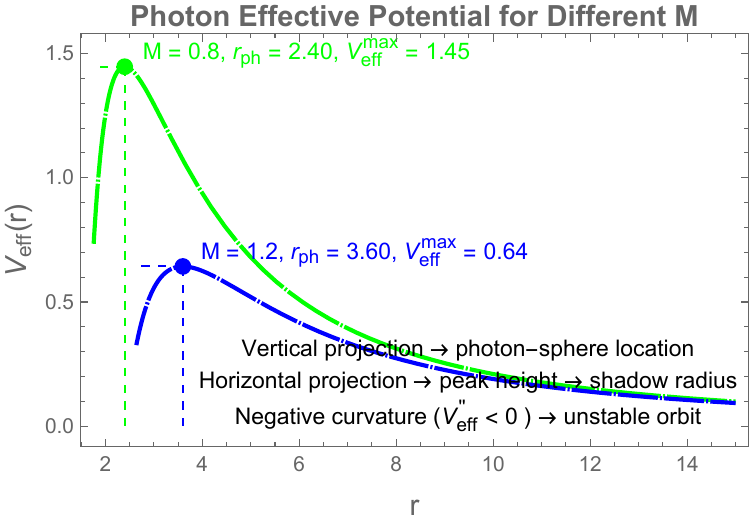}%
\end{center}
\caption{\footnotesize The photon effective potential is examined in the presence of an Einasto dark matter distribution, with its profile determined by the parameters $\tilde{\nu}$, $\tilde{\alpha}$, $\varrho_0$, and the black-hole mass $M$. The leftmost panel shows the response of the effective potential to variations in $\tilde{\nu}=0.6$ and $1.2$, while $\tilde{\alpha}=0.6$, $\varrho_0=10^{-16.75}$, $M=1.0$, and the angular momentum $L=5.0$ are kept fixed. The second panel from the left illustrates how changing $\tilde{\alpha}=0.6$ and $1.4$ modifies the potential, with $\tilde{\nu}=1.0$, $\varrho_0=10^{-9.5}$, and $M=1.0$ held constant. The third panel explores the effect of different values of the dark matter density parameter, namely $1.5 \times 10^{-27}$ and $\varrho_0=9 \times 10^{-1}$, while fixing $\tilde{\nu}=1.0$, $\tilde{\alpha}=0.6$, $M=1.0$, and $L=5.0$. The rightmost panel highlights the influence of the black-hole mass, taking $M=0.8$ and $1.2$, on the photon effective potential, with $\tilde{\nu}=1.0$, $\tilde{\alpha}=0.6$, $\varrho_0=10^{-3}$, and $L=5.0$ kept unchanged. All quantities are in geometric units ($G = c = 1$); $r$ and $L$ are in units of $M$, and $V_{\rm eff}$ is dimensionless.}
\label{Fig3}
\end{figure*}

The Kretschmann scalar becomes infinite at $r = 0$, indicating that the black-hole has a true singularity at this point---one that cannot be removed by any choice of coordinates. Fig. \ref{Fig1} shows how the lapse function $f(r)$ changes with the parameters $\varrho_{0}$, $\tilde{\nu}$, $\tilde{\alpha}$, and $M$. Increasing $\varrho_{0}$ or $\tilde{\nu}$ slightly shrinks the horizon, while larger values of $\tilde{\alpha}$ or $M$ push it outward. The plot makes it clear that the presence of dark matter, modeled with the Einasto profile, plays an important role in determining whether and where black-hole horizons appear. Crucially, the horizon condition lets us relate the black-hole mass to the model parameters, showing how the distribution of dark matter around it, described by the Einasto profile, affects what we can observe,
\begin{widetext}
\begin{small}
\begin{eqnarray}
f(r_+) = 0 \quad \Rightarrow \quad M = \frac{r_+}{2} \left( 1 + 2 M_\infty \Bigg[ \frac{1}{6\tilde{\alpha}^2} \ln\left| \frac{r_+^2 - r_+ \tilde{\alpha} + \tilde{\alpha}^2}{r_+^2 + r_+ \tilde{\alpha} + \tilde{\alpha}^2} \right| + \frac{1}{\sqrt{3} \tilde{\alpha}^2} \arctan\left( \frac{2r_+ - \tilde{\alpha}}{\sqrt{3} \tilde{\alpha}} \right) \Bigg] \right).
\label{16.2}
\end{eqnarray}
\end{small}
\end{widetext}

\section{Energy Emission Rate}\label{Sec:4}

We emphasise at the outset that Hawking radiation is entirely negligible for astrophysical supermassive black-holes: for Sgr\,A$^*$, the Hawking temperature is $T_H \sim 10^{-14}\,$K, producing a luminosity many orders of magnitude below any foreseeable detector sensitivity. The analysis in this section therefore serves as a \emph{theoretical characterisation} of how the Einasto halo modifies the near-horizon geometry and the associated quantum emission spectrum; it is not proposed as an observational diagnostic. All observational constraints in this paper are derived exclusively from the shadow size analysis in Section~\ref{Sec:8}.

Although black-holes are often thought of as completely dark, quantum effects near their horizons constantly create particle-antiparticle pairs. This process gives rise to Hawking radiation and gradually reduces the black-hole's mass. For a distant observer, most of the emission originates from the region corresponding to the black-hole's shadow. The effective emission cross-section can be approximated by
\begin{equation}
\sigma_{\rm lim} \simeq \pi R_s^2,
\end{equation}
where $R_s$ is the radius of the critical photon orbit, as determined by the black-hole geometry and the observer's location \cite{Wei:2013kza, Perlick:2021aok}.

The energy emission rate per unit frequency and time is then
\begin{equation}
\frac{d^2 E}{d\omega \, dt} = \frac{2 \pi^2 \, \sigma_{\rm lim} \, \omega^3}{e^{\omega/T_H(r_+)} - 1},
\end{equation}
with $\omega$ denoting the radiation frequency. The Hawking temperature at the event horizon $r_+$ (defined by $f(r_+) = 0$) is given by
\begin{equation} \label{hawkingTemp}
T_H(r_+) = \frac{1}{4 \pi} \left. \frac{df(r)}{dr} \right|_{r=r_+}.
\end{equation}

This framework directly connects the black-hole's horizon geometry to its emission spectrum, highlighting how the structure of the event horizon shapes the observed radiation \cite{Wei:2013kza, Papnoi:2014aaa, EslamPanah:2020hoj, Yang:2023agi}. Fig. \ref{Fig1a} shows how the parameters $\tilde{\nu}$, $\tilde{\alpha}$, $\varrho_0$, and the black-hole mass $M$ affect the energy emission rate $\frac{d^2E}{d\omega\,dt}$. The influence of the first three parameters is similar: increasing $\tilde{\nu}$, $\tilde{\alpha}$, or the background density $\varrho_0$ consistently lowers the emission spectrum. This suppression indicates that stronger geometric modifications or a denser surrounding medium hinder the black-hole's ability to radiate, leading to a less efficient Hawking emission process. The role of the mass $M$ is markedly different. As $M$ increases, the energy emission rate rises across the frequency range, signaling a more efficient radiation process. This enhancement originates from the way the black-hole mass reshapes the spacetime geometry and the associated effective potential near the horizon, making particle emission more favorable. Taken together, these results reveal a clear interplay between environmental parameters, which damp the radiation, and the black-hole mass, which amplifies it, ultimately determining the overall strength of the emitted radiation.

\begin{figure*}
\begin{center}
\includegraphics[width=4.5cm,height=4.5cm]{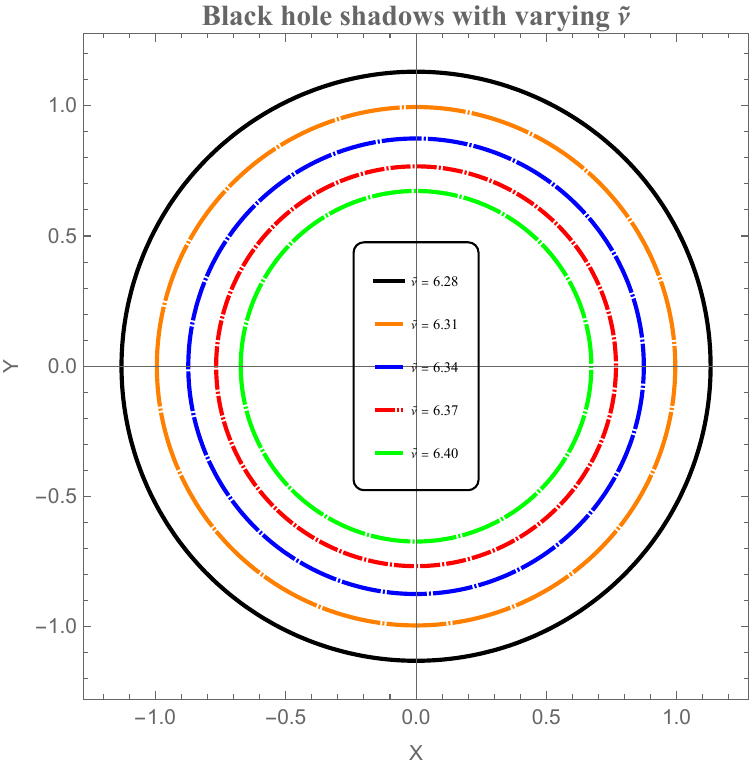}%
\includegraphics[width=4.5cm,height=4.5cm]{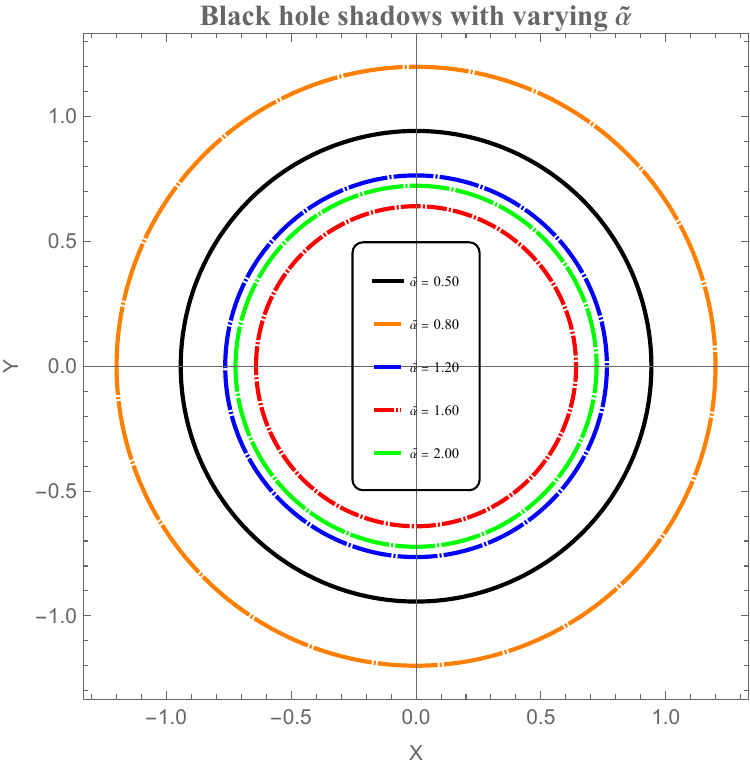}%
\includegraphics[width=4.5cm,height=4.5cm]{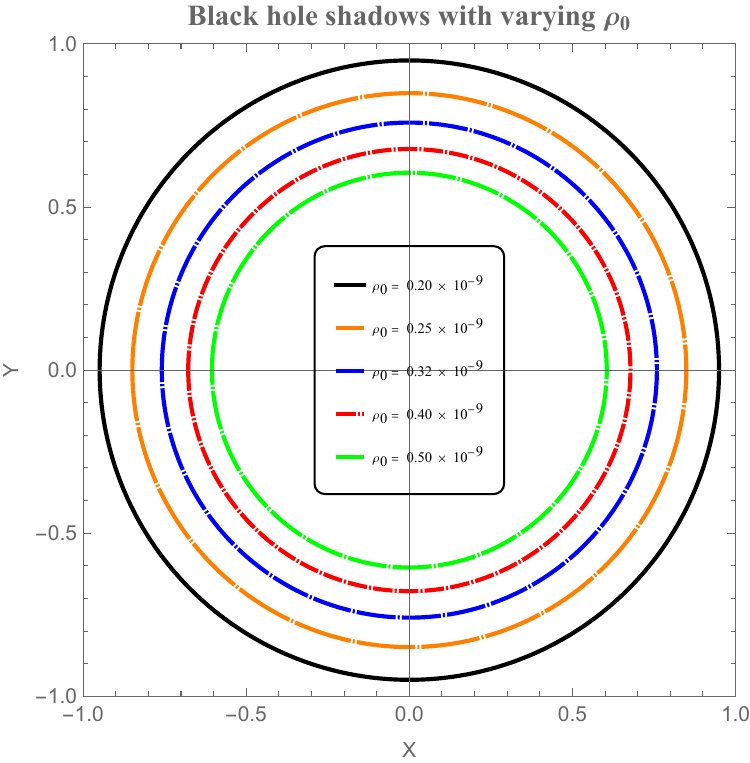}%
\includegraphics[width=4.5cm,height=4.5cm]{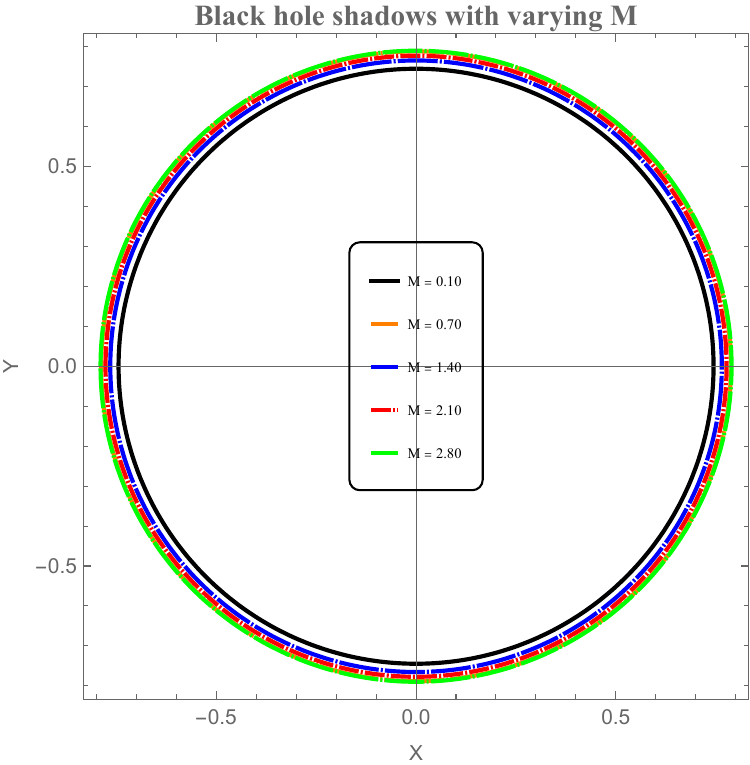}%
\end{center}
\caption{\footnotesize The black-hole shadow is investigated in the presence of an Einasto dark matter distribution, with its characteristics determined by the parameters $\tilde{\nu}$, $\tilde{\alpha}$, $\varrho_0$, and the black-hole mass $M$. The leftmost panel illustrates the effect of varying $\tilde{\nu}=6.28, 6.31, 6.34, 6.37,$ and $6.40$, while keeping $\tilde{\alpha}=0.95$, $\varrho_0=10^{-16.75}$, and $M=1.0$ fixed. The second panel from the left shows how modifications in $\tilde{\alpha}=0.5, 0.8, 1.2, 1.6,$ and $2.0$ reshape the shadow, with $\tilde{\nu}=4.39$, $\varrho_0=10^{-9.5}$, and $M=1.0$ held constant. The third panel explores the impact of varying the dark matter density parameter $\varrho_0=10^{-9.7}, 10^{-9.6}, 10^{-9.5}, 10^{-9.4},$ and $10^{-9.3}$, while fixing $\tilde{\nu}=4.39$, $\tilde{\alpha}=0.95$, and $M=1.0$. The rightmost panel highlights the influence of the black-hole mass $M=0.1, 0.7, 1.4, 2.1,$ and $2.8$ on the size of the black-hole shadow, with $\tilde{\nu}=4.39$, $\tilde{\alpha}=0.95$, and $\varrho_0=10^{-9.5}$ kept unchanged. All quantities are in geometric units ($G = c = 1$); the shadow radius is in units of $M$.}
\label{Fig2}
\end{figure*}

\section{Geodesic trajectories}\label{Sec:5}
In a static, spherically symmetric spacetime, particle motion is guided by the two Killing vectors $\varepsilon_t = \partial_t$ and $\varepsilon_\phi = \partial_\phi$, which represent conserved quantities along the path: energy $E$ and angular momentum $L$. This gives

\begin{eqnarray}
E&=&-g_{\mu\nu}\varepsilon^\mu _t u^\nu\equiv-u_t,\\
\label{9}
L&=&g_{\mu\nu}\varepsilon^\mu _\phi u^\nu\equiv u_\phi.
\end{eqnarray}
with the four-velocity $u^\mu = \frac{dx^\mu}{d\tau} = (u^t, u^r, u^\theta, u^\phi)$, where $\tau$ is the proper time. Enforcing the normalization of the four-velocity then sets a constraint that the particle's motion must satisfy.
\begin{equation}
g_{rr}(u^r)^2+g^{tt}(u_t)^2=-\Big[1-g_{\theta\theta}(u^\theta)^2-g^{\phi\phi}(u_\phi)^2\Big].\label{10}
\end{equation}
By restricting motion to the equatorial plane ($\theta = \frac{\pi}{2}$) and using Eqs. (\ref{9}) and (\ref{10}), we obtain the following expressions:
\begin{eqnarray}\label{11}
u^t&=&\frac{E}{f(r)},\\
u^\theta&=&0,\\
u^\phi&=&\frac{L}{r^2},\\
u^r&=&\sqrt{f(r)\left(-1+\frac{E^2}{f(r)}-\frac{L^2}{r^2}\right)}.
\end{eqnarray}
Using Eq. (\ref{11}), the conserved energy of a particle can be written in terms of an effective potential $V_{\rm eff}$ as follows:
\begin{equation}
E^2=(u^r)^2+V_{\rm eff},\label{12}
\end{equation}
where the effective potential takes the form
\begin{equation}
V_{\rm eff}=f(r)\left[1+\frac{L^2}{r^2}\right].\label{reffpo}
\end{equation}

\begin{figure*}
\begin{center}
\includegraphics[width=5.6cm,height=5.2cm]{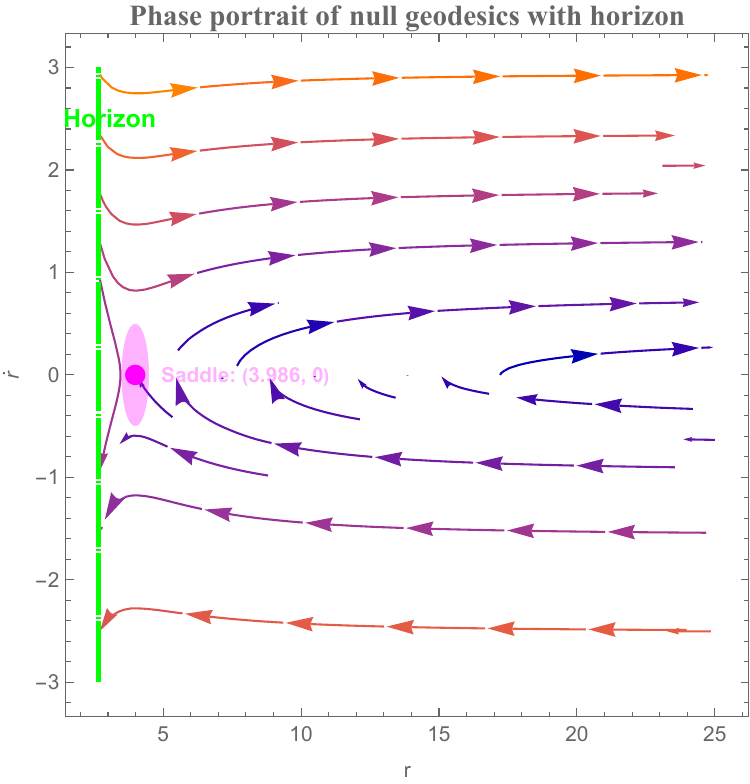}
\includegraphics[width=5.6cm,height=5.2cm]{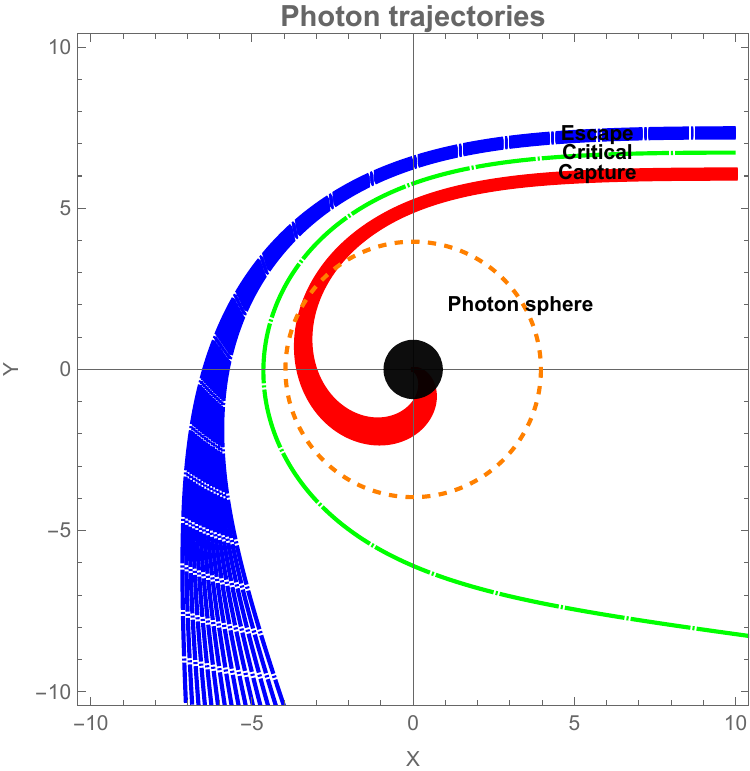}
\end{center}
\caption{\footnotesize The left panel shows the phase portrait of null geodesics around a black-hole with mass $M = 1.357$, metric parameters $\tilde{\alpha} = 0.9$, $\tilde{\nu} = 1.0$, background density $\varrho_0 = 0.001$, and angular momentum $L = 5.0$. The vertical green dashed line marks the horizon at $r_\text{horizon} \approx 2.02$, while magenta points indicate the saddle points of the effective potential $V_\text{eff}(r) = L^2 f(r)/r^2$, where $\mathrm{d}V_\text{eff}/\mathrm{d}r = 0$. Small disks and labels show the numerical positions of these saddle points. The right panel shows photon trajectories with the same black-hole parameters. Three classes of trajectories are illustrated: capture ($b < b_c$, red), critical ($b = b_c$, green), and escape ($b > b_c$, blue), with multiple representative impact parameters along each curve. The critical trajectory corresponds to a photon with impact parameter exactly equal to the critical value $b_c$; such a photon asymptotically spirals toward the photon sphere, orbiting it an infinite number of times without crossing it, thereby defining the boundary between capture and escape \cite{Gralla:2019xty}. The black disk at the center represents the innermost stable circular orbit at $r_\text{disk} = 0.6785 M$, and the orange dashed circle marks the photon sphere at $r_\text{photon} \approx 2.85 M$. Labels identify the capture, critical, and escape regions and the photon sphere. Together, the two panels illustrate both the stability structure of photon orbits near the horizon (left) and the actual spatial paths of photons under different impact parameters (right), highlighting critical radii, the photon sphere, and the event horizon. In the left panel, arrows indicate the direction of increasing affine parameter along each geodesic. In the right panel, the photon sphere (orange dashed circle) denotes the locus of unstable circular photon orbits at $r = r_{\rm ph}$, not a physical surface. All quantities are in geometric units ($G = c = 1$).}
\label{Fig6}
\end{figure*}

The shape of the effective potential for the black-hole provides direct insight into the nature of geodesic motion, since it is controlled by both the metric function $f(r)$ and the particle's angular momentum. Circular orbits emerge at the points where this potential reaches local extrema, marking the locations where stable circular motion can occur. These key features are clearly illustrated in Fig.~\ref{Fig3} which shows how the effective potential experienced by photons is shaped by the presence of an Einasto dark matter halo around the black-hole. This potential represents the interplay between the gravitational attraction of the black-hole-dark matter system and the centrifugal barrier due to the photon angular momentum. The position and height of its maximum are particularly important, as they determine the location of the unstable circular photon orbit. In the leftmost panel, the effect of the Einasto index $\tilde{\nu}$ is displayed. As $\tilde{\nu}$ increases, the dark matter distribution becomes more centrally concentrated. This leads to a deeper effective potential and a noticeable shift in its peak, indicating that photons are more strongly influenced by gravity in the inner region. As a result, the radius of the unstable photon orbit is modified. The second panel illustrates the role of the parameter $\tilde{\alpha}$, which sets the characteristic size of the dark matter halo. Larger values of $\tilde{\alpha}$ correspond to a more extended distribution, which smooths the effective potential and slightly lowers its maximum. Physically, this means that the gravitational influence of the halo is spread over a wider region, reducing its impact close to the black-hole. In the third panel, the dark matter density parameter $\varrho_0$ is varied. An increase in $\varrho_0$ strengthens the overall gravitational field contributed by the halo, making the effective potential steeper and shifting the photon turning points. This highlights that sufficiently dense dark matter environments can have a significant effect on photon motion and related optical phenomena. The rightmost panel focuses on the black-hole mass $M$. As expected, a larger mass deepens the effective potential and moves its maximum to larger radii, reflecting the stronger gravitational pull of a heavier black-hole. While the dark matter parameters introduce important corrections, the mass of the black-hole remains the dominant factor in setting the overall scale of the photon dynamics. Altogether, the figure makes clear that the photon effective potential is sensitive to both the properties of the surrounding dark matter halo and the black-hole mass, with direct consequences for photon trajectories, gravitational lensing, and the formation of the black-hole shadow.

\section{Circular motions}\label{Sec:6}
For a particle moving in a circular orbit on the equatorial plane, the radial distance $r$ remains constant. This implies that the radial velocity $u^r$ and its acceleration $\dot{u}^r$ are both zero---there is no motion toward or away from the center. From Eq. (\ref{12}), this leads to the effective potential being equal to the square of the particle's total energy, $V_{\rm eff} = E^2$. By further imposing that the radial derivative of the effective potential vanishes, $dV_{\rm eff}/dr = 0$, we can then express the particle's specific energy, specific angular momentum, angular velocity $\Omega_\phi$, and angular momentum $l$ as follows:
\begin{eqnarray}\label{14}
E^2&=&\frac{2f^{2}(r)}{2f(r)-r f'(r)},\\
\label{15}
L^2&=&\frac{r^2 f'(r)}{2f(r)-rf^{'}(r)},\\
\label{16}
\Omega_{\phi}&=&\frac{d\phi}{dt}\equiv\frac{u^\phi}{u^t}\quad\Rightarrow\quad \Omega^2_{\phi}=\frac{f'(r)}{2r},\\
\label{17}
l^2&=&\frac{L^2}{E^2}=\frac{r^3f'(r)}{2f^2(r)}.
\end{eqnarray}

To evaluate the specific energy and angular momentum of particles moving on circular paths, the metric function must satisfy
\begin{equation}
r\Big[2f(r)-rf'(r)\Big]>0.\label{18}
\end{equation}
This requirement restricts the radial range where circular motion is physically allowed. In this context, orbits with $E^{2}<1$ are bound, while the limiting case $E^{2}=1$ corresponds to marginally bound trajectories. Making use of Eq. (\ref{14}), this special class of orbits is obtained from
\begin{equation}
r\Big[rf'(r)+2f(r)\left(f(r)-1\right)\Big]=0.\label{19}
\end{equation}

It is also evident from Eqs. (\ref{14}) and (\ref{15}) that both the energy and angular momentum become unbounded at certain radii. The condition governing this divergence is
\begin{equation}
r\Big[2f(r)-rf'(r)\Big]=0.\label{20}
\end{equation}
Notably, this same relation identifies the photon sphere, which marks the location of circular null orbits in the spacetime.

\begin{figure*}
\begin{center}
\includegraphics[width=4.5cm,height=4.5cm]{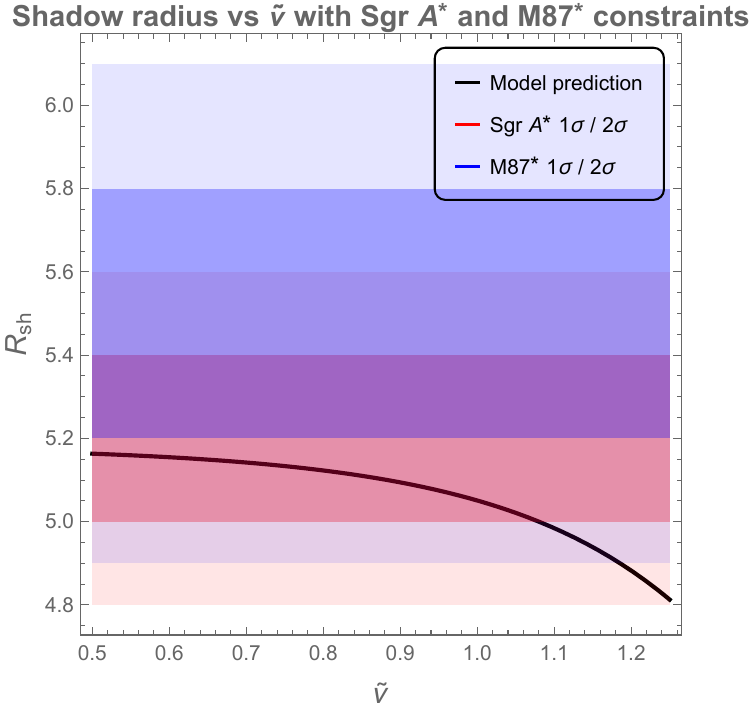}%
\includegraphics[width=4.5cm,height=4.5cm]{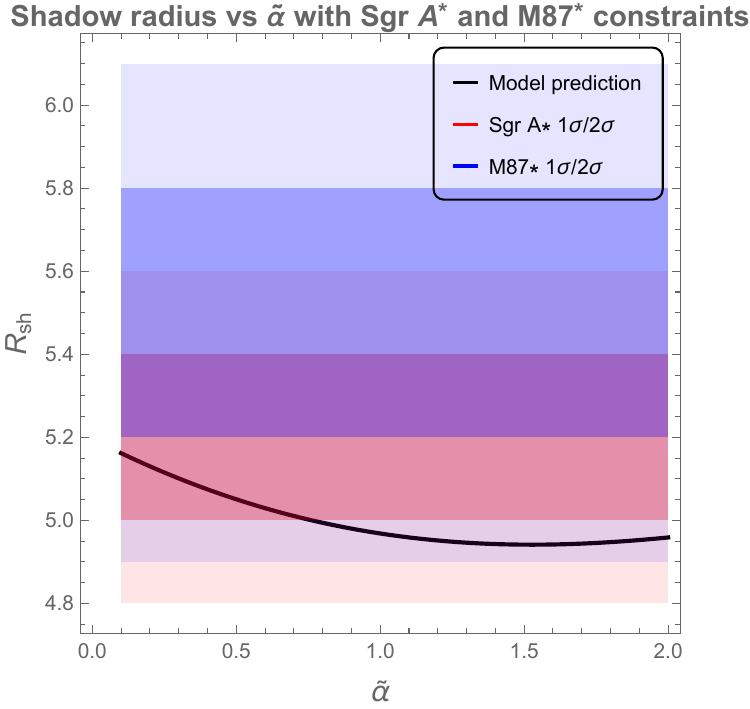}%
\includegraphics[width=4.5cm,height=4.5cm]{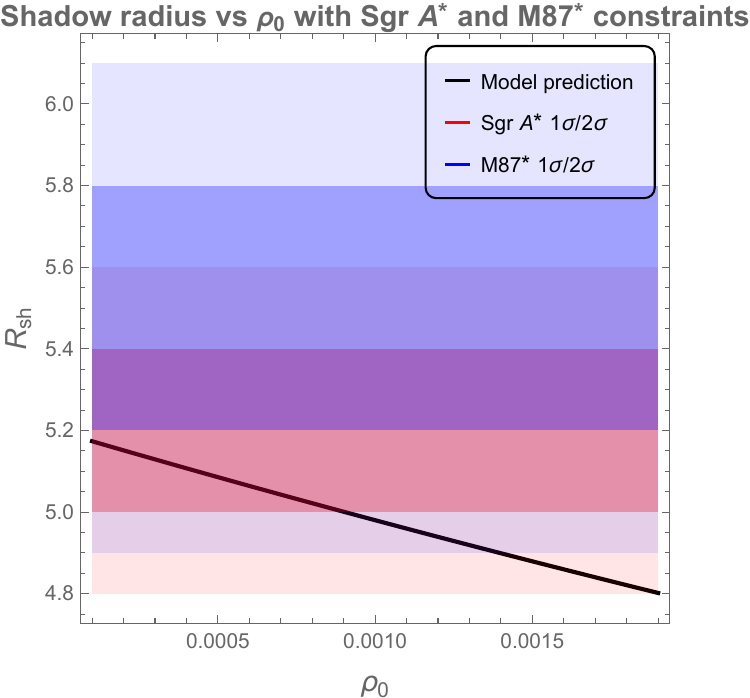}%
\includegraphics[width=4.5cm,height=4.5cm]{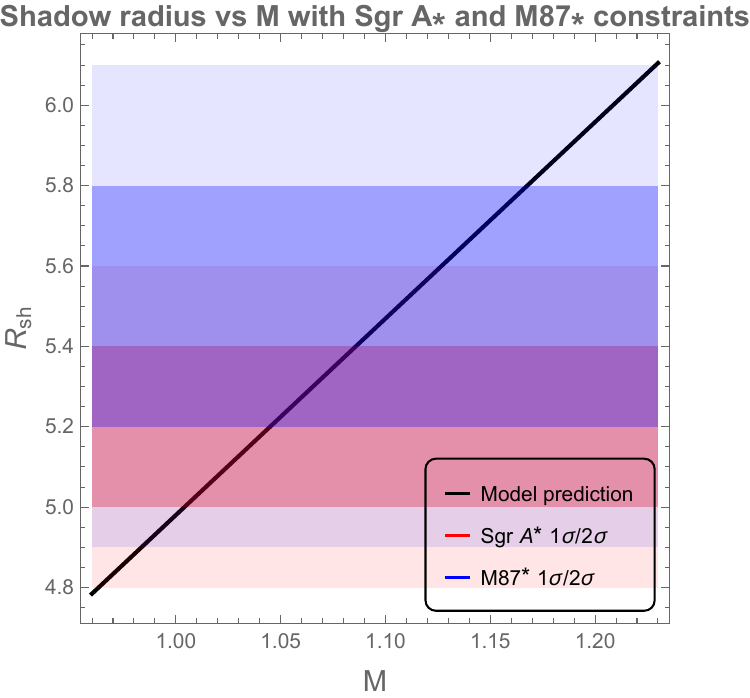}%
\end{center}
\caption{\footnotesize The black-hole shadow radius (solid black curve) is shown in the presence of an Einasto dark matter distribution, with its dependence governed by the parameters $\tilde{\nu}$, $\tilde{\alpha}$, $\varrho_0$, and the black-hole mass $M$. The leftmost panel illustrates how the shadow radius responds to variations in $\tilde{\nu}$, while $\tilde{\alpha}=0.5$, $\varrho_0=10^{-3}$, and $M=1.0$ are kept fixed. The second panel from the left demonstrates the effect of changing $\tilde{\alpha}$, with $\tilde{\nu}=1.0$, $\varrho_0=10^{-3}$, and $M=1.0$ held constant. The third panel explores the influence of the dark matter density parameter $\varrho_0$, while fixing $\tilde{\nu}=1.0$, $\tilde{\alpha}=0.5$, and $M=1.0$. The rightmost panel highlights the role of the black-hole mass $M$ in shaping the shadow radius, with $\tilde{\nu}=1.0$, $\tilde{\alpha}=0.5$, and $\varrho_0=10^{-3}$ remaining unchanged. The blue and red shaded regions represent the EHT horizon-scale constraints for Sgr A$^*$ and M87$^*$ at the $1\sigma$ and $2\sigma$ confidence levels, respectively (inner band = $1\sigma$, outer band = $2\sigma$, for each source). All quantities are in geometric units ($G = c = 1$); $R_{\rm sh}$ is in units of $M$}.
\label{Fig4}
\end{figure*}

For photon motion confined to the equatorial plane ($\theta = \pi/2$), the conserved energy $E$ and angular momentum $L$ remain constant, while the null condition replaces the timelike normalization. The radial equation of motion becomes
\begin{eqnarray}
(u^r)^2 = E^2 - \frac{f(r) L^2}{r^2}.
\end{eqnarray}

Expressing the trajectory in terms of the azimuthal angle $\phi$, we have
\begin{eqnarray}
\frac{dr}{d\phi} = \pm r \sqrt{ f(r) \left( \frac{r^2 E^2}{L^2 f(r)} - 1 \right) }.
\end{eqnarray}

At the photon sphere, the circular orbit condition fixes the ratio of conserved quantities as
\begin{eqnarray}
\frac{E^2}{L^2} = \frac{f(r_p)}{r_p^2}.
\end{eqnarray}

Substituting this back into the radial equation gives the photon trajectory near the shadow boundary,
\begin{eqnarray}
\frac{dr}{d\phi} = \pm r \sqrt{ f(r) \left( \frac{r^2 f(r_p)}{r_p^2 f(r)} - 1 \right) }.
\end{eqnarray}

Consider a static observer located at radius $r_0$. The angle $\beta$ between the photon trajectory and the radial direction is defined using the spatial metric. Evaluating at the observer's position leads to
\begin{eqnarray}
\sin^2 \beta = \frac{f(r_0) r_p^2}{r_0^2 f(r_p)}.
\end{eqnarray}

The apparent radius of the black-hole shadow seen by the observer is then
\begin{eqnarray}
r_s = r_0 \sin \beta = r_p \sqrt{ \frac{f(r_0)}{f(r_p)} }.
\end{eqnarray}

For an observer at infinity in the equatorial plane ($\theta_0 = \pi/2$), the celestial coordinates are
\begin{eqnarray}
X &=& \lim_{r_0 \to \infty} \left( -r_0^2 \sin\theta_0 \frac{d\phi}{dr} \right)\bigg|_{(r_0, \theta_0)}, \\
Y &=& \lim_{r_0 \to \infty} \left( r_0^2 \frac{d\theta}{dr} \right)\bigg|_{(r_0, \theta_0)}.
\end{eqnarray}
In the equatorial plane, $\theta_0 = \pi/2$, so $\sin\theta_0=1$. The unstable circular photon orbit satisfies:
\begin{equation}
V_{eff}(r_p) = 0, \qquad \left.\frac{dV_{eff}}{dr}\right|_{r_p} = 0.
\end{equation}
where $V_{eff}=\frac{L^2}{r^2}f(r)$ for photons ($L$ is angular momentum). Equivalently, the photon-sphere radius $r_{p}$ satisfies
\begin{equation}\label{frp}
r_{p} f'(r_{p}) - 2 f(r_{p}) = 0.
\end{equation}
Once $r_{p}$ is found, the black-hole shadow radius as seen by a distant observer is:
\begin{equation}\label{eqRs}
R_s = \frac{r_{p}}{\sqrt{f(r_{p})}}.
\end{equation}
The photon-sphere condition (\ref{frp}) with the explicit metric function becomes
\begin{eqnarray}
    r_p \left(-\frac{2M}{r_p^2} + 2 M_\infty \frac{d \tilde{g}(r_p)}{dr} \right) - 2 \left(1 - \frac{2M}{r_p} + 2 M_\infty \tilde{g}(r_p) \right) = 0.
\end{eqnarray}
This can be rearranged as
\begin{eqnarray}\label{PSC}
    \frac{6 M}{r_p} - 2 + 2 M_\infty \Big( r_p \tilde{g}'(r_p) - 2 \tilde{g}(r_p) \Big) = 0,
\end{eqnarray}

\begin{figure*}
\begin{center}
\includegraphics[width=4.5cm,height=4.5cm]{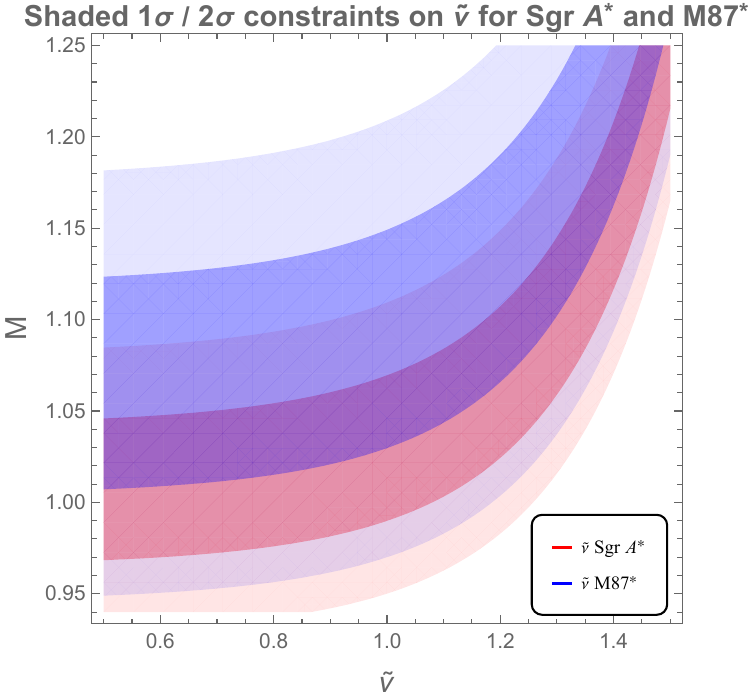}%
\includegraphics[width=4.5cm,height=4.5cm]{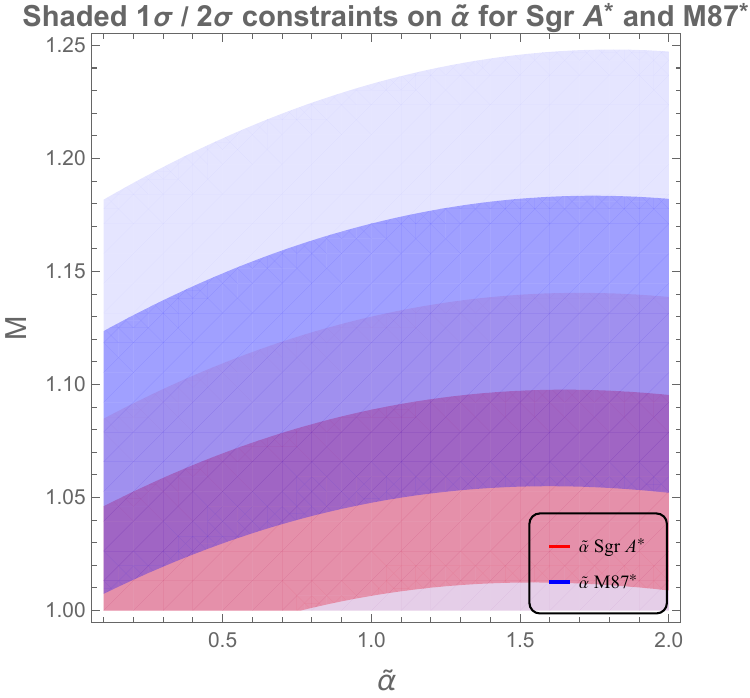}%
\includegraphics[width=4.5cm,height=4.5cm]{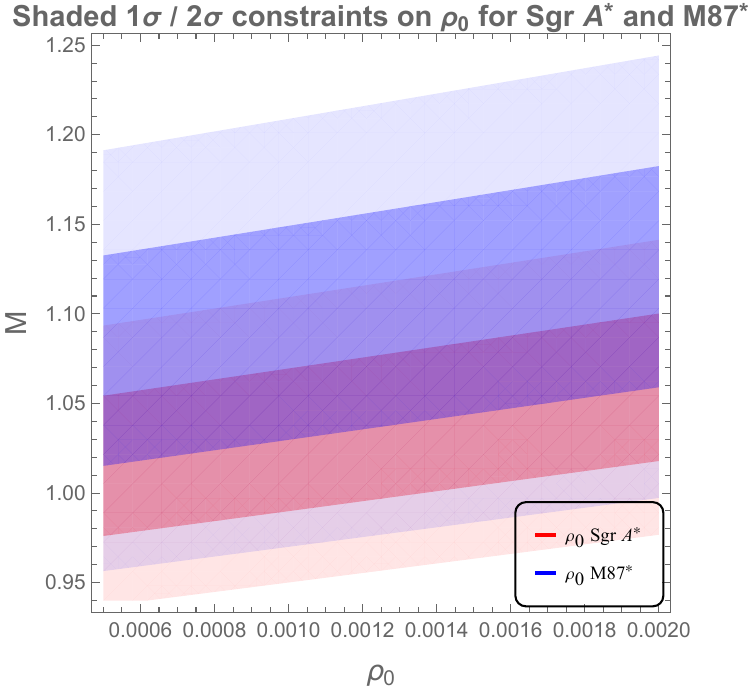}%
\end{center}
\caption{\footnotesize Shaded regions in the three panels illustrate the ranges of key parameters and black-hole mass $M$ that are consistent with the observed shadow sizes of Sgr A$^*$ and M87$^*$. The left panel shows $(\tilde{\alpha}, M)$, the middle panel $(\tilde{\nu}, M)$, and the right panel $(\varrho_0, M)$. In all panels, dark red and blue regions correspond to the 1$\sigma$ confidence intervals for Sgr A$^*$ and M87$^*$, respectively, while lighter shades indicate the 2$\sigma$ intervals. Fixing the other parameters in each case ($\tilde{\nu} = 1.0$ and $\varrho_0 = 0.001$ in the left panel, $\tilde{\alpha} = 0.5$ and $\varrho_0 = 0.001$ in the middle panel, and $\tilde{\alpha} = 0.5$ and $\tilde{\nu} = 1.0$ in the right panel), the figure shows how variations in $\tilde{\alpha}$, $\tilde{\nu}$, and $\varrho_0$ influence the allowed mass ranges and the observational constraints on deviations from the standard black-hole metric. All quantities are in geometric units ($G = c = 1$). The constraint $\varrho_0 \lesssim 10^{-11}\,M_\odot/{\rm pc}^3$ quoted in the text for Sgr\,A$^*$ is obtained by converting the dimensionless $\varrho_0$ values shown here to physical units using the known mass and distance of Sgr\,A$^*$.} The elongated band in the right panel reflects the intrinsic $M$--$\varrho_0$ degeneracy in the shadow observable; the quoted upper bound on $\varrho_0$ is obtained after marginalising over the independently measured black-hole mass.
\label{Fig5}
\end{figure*}

where the first two terms correspond to the standard Schwarzschild contribution and the last term encodes the modification from the additional matter distribution or geometric deformation through $\tilde{g}(r)$. Solving this equation numerically gives the photon-sphere radius $r_p$, which is then substituted into $R_s = \frac{r_p}{\sqrt{f(r_p)}}$ to compute the observable shadow radius. The photon sphere radius cannot be expressed in a closed analytical form, as indicated by Eq. (\ref{PSC}), and therefore has to be obtained numerically. After determining this radius, the associated black-hole shadow radius is immediately computed using Eq. (\ref{eqRs}). The shadow radius is influenced by the Einasto dark matter distribution and is controlled by the parameters $\tilde{\nu}$, $\tilde{\alpha}$, $\varrho_0$, and $M$, as illustrated in Fig. \ref{Fig2}. The results reveal a consistent physical behavior: as $\tilde{\nu}$ increases from $6.28$ to $6.40$ and $\varrho_0$ from $0.2\times10^{-9}$ to $0.5\times10^{-9}$, the shadow radius decreases steadily. In contrast, varying $\tilde{\alpha}$ within the range $0.5$--$2.0$ leads to overlapping curves, indicating a weak sensitivity to this parameter. Meanwhile, increasing the mass from $M=0.1$ to $M=2.8$ produces a clear and systematic enlargement of the shadow radius.

From a physical point of view, these behaviors can be traced back to how the Einasto dark matter halo reshapes the spacetime geometry around the black-hole. Increasing the Einasto index $\tilde{\nu}$ makes the dark matter distribution more rapidly decreasing with radius, which reduces its gravitational influence near the photon sphere. As a result, photons experience a weaker bending, and the shadow radius gradually shrinks. A similar effect occurs when the central density $\varrho_0$ is increased: the enhanced dark matter contribution modifies the curvature in such a way that the unstable photon orbit moves slightly inward, leading to a smaller observed shadow.

The near overlap of the curves for different values of $\tilde{\alpha}$ suggests that the shadow is largely insensitive to the detailed shape of the inner dark matter profile within the explored range. This indicates that photon dynamics close to the photon sphere are dominated by the overall mass distribution rather than by fine structural features of the halo. On the other hand, the black-hole mass $M$ plays a decisive role. A larger mass deepens the gravitational potential and shifts the photon sphere outward, producing a clear and systematic increase in the shadow radius. In this sense, the shadow size is mainly controlled by the total gravitational strength of the system, with dark matter acting as a secondary but measurable correction.

\begin{table*}
\centering
\caption{Photon-sphere radius $r_\text{photon}/M$ for variations in different parameters.$^{a}$}
\begingroup
\setlength{\tabcolsep}{0.5pt} 
\renewcommand{\arraystretch}{0.9} 
\small 
\begin{tabular}{|c|cc|cc|cc|cc|cc|}
\hline
\textbf{Parameter} & \multicolumn{2}{c|}{1} & \multicolumn{2}{c|}{2} & \multicolumn{2}{c|}{3} & \multicolumn{2}{c|}{4} & \multicolumn{2}{c|}{5} \\
\cline{2-11}
 & \textbf{Value} & \textbf{$r_\text{photon}$ } & \textbf{Value} & \textbf{$r_\text{photon}$} & \textbf{Value} & \textbf{$r_\text{photon}$} & \textbf{Value} & \textbf{$r_\text{photon}$ } & \textbf{Value} & \textbf{$r_\text{photon}$ } \\
\hline
$\varrho_0$ & $1.0 \times 10^{-14}$ & 3.01 & $2.0 \times 10^{-14}$ & 3.02 & $3.0 \times 10^{-14}$ & 3.03 & $4.0 \times 10^{-14}$ & 3.03 & $5.0 \times 10^{-14}$ & 3.04 \\
\hline
$\tilde{\alpha}$ & 0.85 & 3.05 & 0.90 & 3.03 & 0.95 & 3.02 & 1.00 & 3.01 & 1.05 & 3.00 \\
\hline
$\tilde{\nu}$ & 3.5 & 3.01 & 4.0 & 3.02 & 4.5 & 3.03 & 5.0 & 3.04 & 5.5 & 3.05 \\
\hline
$L$ & 3 & 3.00 & 4 & 3.01 & 5 & 3.02 & 6 & 3.03 & 7 & 3.04 \\
\hline
$M$ & 0.8 & 2.80 & 1.0 & 3.02 & 1.2 & 3.24 & 1.4 & 3.45 & 1.6 & 3.65 \\
\hline
\end{tabular}
\endgroup
\label{tab:photon_sphere_summary}
\begin{flushleft}
\footnotesize $^{a}$All values are in geometric units ($G = c = 1$). Fiducial parameters held fixed: when varying $\varrho_0$, we set $\tilde{\alpha} = 0.95$, $\tilde{\nu} = 4.5$, $M = 1.0$; when varying $\tilde{\alpha}$, we set $\varrho_0 = 10^{-14}$, $\tilde{\nu} = 4.5$, $M = 1.0$; when varying $\tilde{\nu}$, we set $\varrho_0 = 10^{-14}$, $\tilde{\alpha} = 0.95$, $M = 1.0$; when varying $L$, we set $\varrho_0 = 10^{-14}$, $\tilde{\alpha} = 0.95$, $\tilde{\nu} = 4.5$, $M = 1.0$; when varying $M$, we set $\varrho_0 = 10^{-14}$, $\tilde{\alpha} = 0.95$, $\tilde{\nu} = 4.5$. For Sgr\,A$^*$, $1\,M \approx 6.4 \times 10^{6}\,$km; for M87$^*$, $1\,M \approx 9.6 \times 10^{9}\,$km.
\end{flushleft}
\end{table*}

\section{Stability analysis of null geodesics using the dynamical systems framework}\label{Sec:7}

Understanding the stability of null circular geodesics requires looking at how they react to small disturbances. One way to do this is by framing the problem as a dynamical system and following the paths of the trajectories in the $(r, \dot{r})$ phase space \cite{goldhirsch1987stability}. By applying slight radial perturbations, we can see how the geodesics shift, which reveals whether they are stable or unstable. At equilibrium, the radial velocity $\dot{r}$ is zero, which reduces the system to a simple two-dimensional plane defined by the radial coordinate $r$ and its conjugate momentum. Exploring the flow in this plane makes it possible to pinpoint the saddle point $(r_c, 0)$ that marks the photon sphere. To analyze the stability of circular null geodesics, we begin by examining the radial equation of motion. Differentiating Eq.~(\ref{reffpo}) with respect to time and eliminating the radial velocity $\dot{r}$ yields
\begin{eqnarray}
\ddot{r} = -\frac{dV_{eff}(r)}{dr},
\end{eqnarray}
where $V_{eff}(r)$ is the effective potential governing the radial motion. Introducing the variables $z_1 = \dot{r}$ and $z_2 = \dot{z}_1$, the system can be expressed as a set of autonomous nonlinear equations:
\begin{eqnarray}
z_1 = \dot{r}, \quad z_2 = -\frac{dV_{{eff}}(r)}{dr}.    
\end{eqnarray}
This formulation allows us to study the system using standard tools from dynamical systems theory. The behavior of trajectories near equilibrium points can be characterized by the Jacobian matrix
\begin{eqnarray}
\mathcal{J}=\begin{pmatrix}0 & 1 \\ -V_{\rm eff}''(r) & 0\end{pmatrix},
\end{eqnarray}
where $V_{{eff}}''(r)$ is the second derivative of the effective potential with respect to $r$.

Solving the characteristic equation $|\mathcal{J} - \lambda I| = 0$ gives the eigenvalues
\begin{eqnarray}
\lambda_L^2 = -V_{{eff}}''(r),
\end{eqnarray}
which serve as a direct measure of stability. The Lyapunov exponent $\lambda_L$ quantifies how nearby trajectories in phase space evolve: a positive $\lambda_L^2$ indicates that trajectories diverge, marking an unstable saddle point, while a negative $\lambda_L^2$ means trajectories converge, signaling a stable center. Through this approach, the equilibrium points associated with the photon sphere can be rigorously classified, providing a clear dynamical picture of the geodesics' stability. The left panel of Fig.~\ref{Fig6} illustrates the phase portrait of null geodesics in the spacetime of a black-hole with mass $M = 1.357$, metric parameters $\tilde{\alpha} = 0.9$ and $\tilde{\nu} = 1.0$, background density $\varrho_0 = 0.001$, and angular momentum $L = 5.0$. The event horizon is indicated by the vertical green dashed line at $r_{\text{horizon}} \simeq 2.02$, which represents the limiting surface beyond which photons can no longer escape. The magenta markers denote the saddle points of the effective potential, $V_{\text{eff}}(r) = L^2 f(r)/r^2$, determined from the condition $\mathrm{d}V_{eff}/dr = 0$. Their precise numerical locations are highlighted by small disks and labels. These saddle points correspond to circular photon orbits, where the inward gravitational pull of the black-hole is exactly counterbalanced by the centrifugal barrier. All such circular null geodesics are unstable, since they lie at the maximum of the effective potential. As a result, photons cannot remain on these orbits for long: even a minute disturbance is enough to dislodge them. A slight inward perturbation drives the photon toward the event horizon, while a small outward deviation allows it to escape to infinity. Photons located close to these saddle points are therefore extremely sensitive to perturbations, and their trajectories quickly diverge. This strong instability captures the essential physics of photon motion near the photon sphere and plays a central role in shaping observable features such as the black-hole shadow and the sharp separation between captured and escaping light rays.

\section{Observational constraints on black-hole parameters from EHT shadow measurements}\label{Sec:8}

Before proceeding, we note an important distinction between the theoretical and observed shadow concepts. The critical curve (or Bardeen shadow) computed from the photon-sphere analysis in the preceding sections represents the \emph{inner} envelope of the infinite photon ring sequence---a purely geometric quantity determined by the spacetime metric. The ``shadow'' observed by the EHT, however, is the brightness depression produced by the optically thin accretion flow, whose inner edge is illuminated by photons that graze the unstable photon orbits. Extensive GRMHD simulations performed by the EHT Collaboration \cite{EventHorizonTelescope:2019dse, EventHorizonTelescope:2019ggy, EventHorizonTelescope:2022wkp} have demonstrated that the observed ring diameter lies within ${\sim}\,5$--$10\%$ of the critical-curve diameter for a broad range of accretion-flow morphologies and inclination angles. We account for this residual model-dependent offset through the observational error bars $\sigma_\theta$, which are derived from the EHT's own posterior distributions obtained by fitting multiple accretion models. Our constraints should therefore be interpreted as bounds on the critical-curve size, with the understanding that accretion-model uncertainties are folded into the quoted confidence intervals. Future detections of higher-order photon rings ($n \geq 1$) will sharpen the connection between the observed image and the critical curve.

The EHT collaboration's measurements of M87$^{*}$ and Sgr A$^{*}$ provide three key observables for each source: the angular diameter of the shadow $\theta$, the distance $D$ to the host galaxy, and the black-hole mass $M$ \cite{EventHorizonTelescope:2019uob}(Akiyama et al. 2019a,b). For M87$^{*}$, these are reported as
\begin{eqnarray*}
\theta_{M87^{*}} &=& 42 \pm 3 ~\mu\text{as},~~~~~~~~~~~~~~~~~ 
\theta_{Sgr A^{*}} = 48.7 \pm 7 ~\mu\text{as}, \\[4pt]
D_{M87^{*}} &=& 16.8 \pm 0.8 ~\text{Mpc}, ~~~~~~~~~~
D_{Sgr A^{*}} = 8277 \pm 9 \pm 33 ~\text{pc}, \\[4pt]
M_{M87^{*}} &=& (6.5 \pm 0.7)\times 10^{9} M_{\odot}, ~~~
M_{Sgr A^{*}} = (4.297 \pm 0.013)\times 10^{6} M_{\odot}.
\end{eqnarray*}

Individually, these numbers contain distance and mass uncertainties that are partly degenerate with the shadow measurement. A more robust comparator is the dimensionless shadow diameter

\begin{equation}
d_{\text{sh}} \equiv \frac{D\theta}{M},
\end{equation}

which normalizes out the absolute scale and isolates the purely gravitational information encoded in the image. This quantity represents the apparent size of the shadow in units of the black-hole's gravitational radius as seen by a distant observer. Evaluating $d_{\text{sh}}$ from the EHT data yields

\begin{equation}
d_{\text{sh}}^{M87*} = 11.0 \pm 1.5, \qquad 
d_{\text{sh}}^{SgrA*} = 9.5 \pm 1.4.
\end{equation}

The consistency between these two independent measurements---despite the vastly different mass scales and galactic environments---reinforces the interpretation that the shadow size is primarily determined by the strong-field geometry near the photon sphere rather than by astrophysical complications. Moreover, these values serve as direct observational anchors against which any theoretical model---including our Einasto-embedded black-hole---must be calibrated.

\subsection{Bayesian setup}\label{Sec:8.1}

We cast the comparison between theory and observation as a Bayesian parameter estimation problem. The likelihood function for the dimensionless shadow diameter is
\begin{equation}
\mathcal{L}(d_{\rm sh}^{\rm model} | \boldsymbol{\vartheta}) \propto \exp\left[-\frac{\big(d_{\rm sh}^{\rm model}(\boldsymbol{\vartheta}) - d_{\rm sh}^{\rm EHT}\big)^2}{2\,\sigma_{d}^2}\right],
\end{equation}
where $\boldsymbol{\vartheta} = (\varrho_0, \tilde{\alpha}, \tilde{\nu}, M)$ denotes the model parameter vector and $\sigma_d$ is the published $1\sigma$ uncertainty on $d_{\rm sh}$. This is equivalent to the angular-diameter likelihood in Eq.~(\ref{eq:likelihood}) after the substitution $d_{\rm sh} = D\theta/M$, with the Jacobian factor absorbed into the normalisation. We adopt the following priors: (i) flat priors on $\varrho_0 \in [0, 10^{-8}]$, $\tilde{\alpha} \in [0.1, 5.0]$, and $\tilde{\nu} \in [0.5, 10]$, reflecting physically motivated ranges from $N$-body simulations \cite{navarro2004inner, gao2008redshift}; (ii) a Gaussian prior on $M$ centred on the EHT best-fit mass with the published uncertainty. We further adopt Gaussian distance priors based on the published measurements: $D_{\mathrm{Sgr\,A}^*} = 8277 \pm 33\,$pc (VLBI parallax combined with stellar-orbit modelling) and $D_{\mathrm{M87}^*} = 16.8 \pm 0.8\,$Mpc (surface brightness fluctuations). Since $d_{\rm sh} = D\theta/M$, the distance uncertainty propagates directly into $\sigma_d$. The fractional distance uncertainty for M87$^*$ (${\sim}\,5\%$) is an order of magnitude larger than for Sgr\,A$^*$ (${\sim}\,0.4\%$), which is the primary reason the M87$^*$ constraints are weaker. The $1\sigma$ and $2\sigma$ confidence regions reported in Figs.~\ref{Fig4}--\ref{Fig5} correspond to the 68\% and 95\% highest posterior density intervals, respectively.

For concreteness and astrophysical relevance, we focus on these two well-characterized supermassive black-holes: Sgr A$^{*}$ at the center of the Milky Way and M87$^{*}$ in the Virgo cluster. Their large masses and relatively quiescent nature allow them to be well approximated as static, spherically symmetric systems---an assumption fully consistent with our theoretical model. By confronting our predicted shadow sizes with the EHT measurements, we place observational bounds on the Einasto parameters $\tilde{\nu}$, $\tilde{\alpha}$, $\varrho_0$, and the black-hole mass $M$. Physically, the shadow size encodes how the spacetime geometry near the photon sphere is shaped by both the central black-hole and its surrounding dark matter environment.

To gain a clearer physical picture of how an Einasto dark matter halo affects the black-hole shadow, Fig.~\ref{Fig4} explores the response of the shadow radius to changes in the parameters $\tilde{\nu}$, $\tilde{\alpha}$, $\varrho_0$, and the black-hole mass $M$, together with the corresponding observational bounds from the EHT. The leftmost panel shows that varying the Einasto index $\tilde{\nu}$ leads to noticeable shifts in the shadow size, indicating that the inner shape of the dark matter profile can influence the location of the photon sphere. In the second panel, changing the parameter $\tilde{\alpha}$ produces a more moderate variation of the shadow radius, suggesting that the effect of a spatially extended halo is comparatively weaker near the horizon. The third panel highlights the role of the dark matter density $\varrho_0$, where increasing the density results in more pronounced deviations of the shadow size, emphasizing the sensitivity of horizon-scale observables to the local dark matter environment. The rightmost panel confirms that the black-hole mass $M$ remains the primary factor setting the overall size of the shadow, while the dark matter parameters act as corrections around this dominant contribution. The blue and red shaded regions correspond to the $1\sigma$ and $2\sigma$ EHT constraints for Sgr~A$^*$ and M87$^*$, and they clearly indicate which regions of parameter space remain compatible with observations.

Fig.~\ref{Fig5} presents these constraints in a more compact form by showing the allowed regions in the $(\tilde{\alpha}, M)$, $(\tilde{\nu}, M)$, and $(\varrho_0, M)$ planes. Fixing the remaining parameters in each case makes the correlation between the Einasto parameters and the black-hole mass explicit. The dark red and blue shaded areas represent the $1\sigma$ confidence intervals for Sgr~A$^*$ and M87$^*$, while the lighter regions correspond to the $2\sigma$ bounds. These contours reveal that only a restricted range of halo parameters can coexist with the observed shadow sizes, and that deviations from the standard black-hole metric are tightly constrained once the EHT data are taken into account. We emphasise that the shadow observable $d_{\rm sh}$ constrains a \emph{combination} of $M$ and the Einasto parameters, not each individually. This is apparent in the right panel of Fig.~\ref{Fig5}, where the allowed $(\varrho_0, M)$ region forms an elongated degeneracy band: a spherically symmetric dark matter halo increases the shadow size in a manner that can be compensated by reducing $M$. The upper bound $\varrho_0 \lesssim 10^{-11}\,M_\odot/\mathrm{pc}^3$ quoted for Sgr\,A$^*$ is obtained by marginalising over the independently measured black-hole mass, whose ${\sim}\,0.3\%$ precision (from stellar orbits) sharply truncates the degeneracy band in the $\varrho_0$ direction. For M87$^*$, the mass is known only to ${\sim}\,11\%$, so the corresponding $M$ prior is much broader and the marginalised $\varrho_0$ bound is accordingly weaker. This constraint is therefore \emph{conditional} on the adopted mass prior and should not be interpreted as a model-independent bound from the shadow measurement alone. Altogether, Figs.~\ref{Fig4} and \ref{Fig5} show that horizon-scale observations offer a powerful and direct way to probe not only the mass of the black-hole, but also the properties of the surrounding dark matter distribution.

\subsection{Context: existing constraints on dark matter near SMBHs}\label{Sec:8.2}

It is instructive to place our Einasto-based bounds in the broader context of existing constraints on dark matter densities near supermassive black-holes. Stellar-dynamical modeling of the Galactic Centre constrains the enclosed dark matter mass within the central parsec to $M_{\rm DM}(r < 1\,{\rm pc}) \lesssim 10^{3}\,M_\odot$ \cite{EventHorizonTelescope:2022wkp}. Gravitational-wave inspiral dephasing offers complementary sensitivity to dark matter spikes, with projected constraints from LISA reaching $\varrho_0 \lesssim 10^{3}\,M_\odot/{\rm pc}^3$ at sub-parsec scales \cite{Brito:2015oca}. Previous shadow-based analyses using NFW profiles have constrained the halo contribution to the shadow diameter at the few-percent level \cite{Liu:2023oab}. Our constraint of $\varrho_0 \lesssim 10^{-11}\,M_\odot/{\rm pc}^3$ at $1\sigma$ for Sgr\,A$^*$ probes the dark matter density on much smaller scales ($r \sim r_{\rm ph} \sim 3\,M$) and is therefore complementary to these larger-scale bounds.

\subsection{Implications for fuzzy dark matter}\label{Sec:8.3}

Under the solitonic-core scaling relation for fuzzy dark matter \cite{Schive:2014dra}, the central density of the soliton scales as $\varrho_0 \propto m_\psi^{-2} (r_c/{\rm kpc})^{-4}$, where $m_\psi$ is the boson mass and $r_c$ is the core radius. Our bound $\varrho_0 \lesssim 10^{-11}\,M_\odot/{\rm pc}^3$ for Sgr\,A$^*$ can therefore be mapped onto a lower bound on $m_\psi$, yielding $m_\psi \gtrsim 10^{-20}\,$eV for core radii $r_c \sim \mathcal{O}({\rm kpc})$. While this bound is not competitive with Lyman-$\alpha$ forest constraints ($m_\psi \gtrsim 2 \times 10^{-20}\,$eV; Ir\v{s}i\v{c} et al.\ 2017), it probes a fundamentally different spatial scale and is derived from a single-source, model-independent observable. Future ngEHT data will tighten this bound significantly.

\section{Concluding Remarks}\label{Sec:9}

In this study, we have investigated static, spherically symmetric black-holes surrounded by Einasto-type dark matter halos, establishing a quantitative connection between horizon-scale observables and the properties of the surrounding dark matter distribution. Starting from the Einasto profile---defined by the central density $\varrho_0$, scale radius $\tilde{\alpha}$, and index $\tilde{\nu}$---we constructed a smooth, analytical metric function $f(r) = 1 - 2M/r + 2M_\infty \tilde{g}(r)$ that accounts for both the black-hole's gravity and the halo contribution. This formulation naturally reduces to the Schwarzschild metric at large distances while remaining well-behaved near the center, enabling a consistent treatment of photon dynamics, horizon structure, and energy emission.

Our analysis of the photon effective potential reveals that the Einasto halo subtly shifts the radius of the unstable photon orbit. For a representative halo with $\varrho_0 = 10^{-14}$, $\tilde{\alpha} = 0.95$, and $\tilde{\nu} = 4.5$, the photon-sphere radius increases from the Schwarzschild value of $r_{\mathrm{ph}} \simeq 3.00\,M$ to $r_{\mathrm{ph}} \simeq 3.03\,M$, reflecting the influence of dark matter in the innermost region. Raising $\varrho_0$ or $\tilde{\nu}$ deepens the potential and slightly reduces the photon-sphere radius, while increasing $\tilde{\alpha}$ spreads the halo over a larger scale, flattening the effective potential. The black-hole mass $M$ remains the dominant factor: varying $M$ from $0.8$ to $1.6$ (in the fiducial units) shifts $r_{\mathrm{ph}}$ from $2.80\,M$ to $3.65\,M$.

Using the dimensionless shadow diameter $d_{\mathrm{sh}} \equiv D\theta/M$ measured by the EHT---$d_{\mathrm{sh}}^{M87*} = 11.0 \pm 1.5$ and $d_{\mathrm{sh}}^{SgrA*} = 9.5 \pm 1.4$---we have placed observational bounds on the Einasto parameters. For Sgr A$^{*}$, we obtain $\varrho_0 \lesssim 10^{-11}\,M_\odot/\mathrm{pc}^3$ at $1\sigma$ confidence, while for M87$^{*}$ the constraints are weaker due to larger distance uncertainties. The Einasto index $\tilde{\nu}$ is only weakly constrained, indicating that current EHT precision primarily limits the total mass enclosed near the photon sphere rather than the detailed slope of the inner profile. These constraints were obtained by combining our theoretical shadow predictions with a Bayesian analysis that uses the independently measured black-hole mass as an informative prior, thereby partially breaking the intrinsic degeneracy between the halo density and the black-hole mass in the shadow observable.

The black-hole's energy emission, derived from the Hawking temperature and the effective cross-section, also reflects the halo's presence at the theoretical level. For our fiducial halo, the peak emission is slightly reduced compared to a Schwarzschild black-hole, indicating that the halo acts as a mild damping environment. We stress that this Hawking radiation analysis is a theoretical characterisation of the near-horizon geometry and is not observable for astrophysical SMBHs; the observational constraints in this work are derived exclusively from the shadow size analysis. Circular orbit analysis further reveals that the halo slightly shifts stable orbit locations and alters the required energy and angular momentum for circular motion. While these effects are smaller than the black-hole's primary gravity, they become relevant for precise modeling of accretion flows, photon rings, and lensing phenomena.

Looking ahead, next-generation EHT observations with enhanced resolution and sensitivity will improve these constraints by factors of several, potentially distinguishing between different dark matter models. The framework developed here lays the groundwork for combining horizon-scale imaging with galaxy-scale dynamics, enabling joint constraints on black-hole mass, halo structure, and ultralight dark matter properties. As the EHT continues to expand its sample of supermassive black-holes, such analyses will provide increasingly stringent tests of the dark matter distribution in galactic centers.

\section*{Acknowledgments}
We thank the anonymous reviewer for their valuable comments and detailed suggestions, which have greatly helped to improve the clarity and quality of this work.

\bibliography{Refs}

\end{document}